\documentclass[preprint,onecolumn,nofootinbib,superscriptaddress,11pt]{revtex4-2}
\usepackage[utf8]{inputenc}
\usepackage{amsmath}
\usepackage{amssymb}
\usepackage{amsfonts}
\usepackage{graphicx}
\usepackage{epstopdf}
\usepackage[caption=false]{subfig}
\usepackage{multirow}
\usepackage{hyperref}
\usepackage{color}

\usepackage{setspace}
\let\oldfootnote\footnote 
\renewcommand{\footnote}[1]{\oldfootnote{\setstretch{1}#1}} 

\usepackage{footmisc}
\usepackage[T1]{fontenc}
\usepackage[utf8]{inputenc}

\usepackage{amsmath}
\usepackage{color}
\usepackage{graphicx}
\usepackage{slashed}
\usepackage{tcolorbox}
\usepackage{ulem}

\usepackage{amsmath}
\usepackage{graphicx,float,color}
\usepackage{hyperref}
\hypersetup{colorlinks = true, linkcolor = blue, urlcolor  = blue, citecolor = blue, anchorcolor = blue}
\usepackage{epic}
\usepackage{eepic}
\usepackage{epsfig}
\usepackage{latexsym}
\usepackage{color}
\usepackage{url}  
\usepackage{float}
\usepackage{bm}
\usepackage{geometry}
\usepackage{slashed}
\usepackage[detect-all]{siunitx}

\usepackage{physics}
\usepackage{amsmath} 
\usepackage{amsfonts}
\usepackage{color}
\usepackage{url}
\usepackage{graphicx,float,color}
\usepackage{verbatim}

\usepackage{subcaption}
\usepackage{mathtools}

\providecommand{\mnras}{ Mon. Not. Roy. Astron. Soc.}

\def \dt           {\delta t}
\def \dPA      {\delta \rm PA}
\def \dpa        {\Delta \theta}

\def \1           { n_{p,n}^{\dpa} }
\def \2          { n_{q,m}^{\dpa} }
\def \3          { n_{r,l}^{\dt} }
\def \4          { n_{p,n'}^{\dpa} }
\def \5          { n_{q,m'}^{\dpa} } 
\def \6          { n_{r,l'}^{\dt} }

\newcommand{\PA}{{\rm PA}}
\newcommand{\bC}{{\boldsymbol{C}}}
\newcommand{\bF}{{\boldsymbol{F}}}

\newcommand{\bphi}{{\boldsymbol{\phi}}}
\newcommand{\mC}{{\mathcal{C}}}
\newcommand{\bX}{{\boldsymbol{X}}}
\newcommand{\bY}{{\boldsymbol{Y}}}
\newcommand{\bD}{{\boldsymbol{D}}}
\newcommand{\bS}{{\boldsymbol{S}}}
\newcommand{\bB}{{\boldsymbol{B}}}
\newcommand{\bV}{{\boldsymbol{V}}}

\newcommand{\bDvth}{{\boldsymbol{\vartheta}}}
\newcommand{\bDPA}{{\boldsymbol{\delta} \mathbf{ PA}}}
\newcommand{\bDt}{{\boldsymbol{\delta t}}}

\makeatletter
\def\ps@pprintTitle{%
 \let\@oddhead\@empty
 \let\@evenhead\@empty
 \def\@oddfoot{}%
 \let\@evenfoot\@oddfoot}
\makeatother

\begin{document}

\title{Probing Ultralight Axion-like Dark Matter: A Pulsar Timing Arrays-Pulsar Polarization Arrays Synergy}

\author{Ximeng Li}
\affiliation{Institute of High Energy Physics, Chinese Academy of Sciences, Beijing 100049, China}
\affiliation{ School of Physics Sciences, University of Chinese Academy of Sciences, Beijing 100039, China}

\author{Yonghao Liu}
\affiliation{Department of Physics and Jockey Club Institute for Advanced Study,  
The Hong Kong University of Science and Technology, Hong Kong S.A.R., China}

\author{Zu-Cheng Chen}
\affiliation{Department of Physics and Synergetic Innovation Center for Quantum Effects and Applications, Hunan Normal University, Changsha, Hunan 410081, China}
\affiliation{Institute of Interdisciplinary Studies, Hunan Normal University, Changsha, Hunan 410081, China}

\author{Shi Dai}
\affiliation{Australia Telescope National Facility, CSIRO, Space and Astronomy, PO Box 76, Epping, NSW 1710, Australia}

\author{Boris Goncharov}
\affiliation{Max Planck Institute for Gravitational Physics (Albert Einstein Institute), 30167 Hannover, Germany}
\affiliation{Leibniz Universit¨at Hannover, 30167 Hannover, Germany}

\author{Xiao-Song Hu}
\affiliation{School of Physics and Astronomy, Beijing Normal University, Beijing 100875, China}
\affiliation{Department of Physics,  Faculty of Arts and Sciences, Beijing Normal University, Zhuhai 519087, China}

\author{Qing-Guo Huang}
\affiliation{ School of Fundamental Physics and Mathematical Sciences, Hangzhou Institute for Advanced Study, UCAS, Hangzhou 310024, China}
\affiliation{  School of Physical Sciences, University of Chinese Academy of Sciences, No. 19A Yuquan Road, Beijing 100049, China}
\affiliation{Institute of Theoretical Physics, Chinese Academy of Sciences, Beijing 100190, China}

\author{Tao Liu}
\email{taoliu@ust.hk}
\affiliation{Department of Physics and Jockey Club Institute for Advanced Study,  
The Hong Kong University of Science and Technology, Hong Kong S.A.R., China}

\author{Jing Ren}
\email{renjing@ihep.ac.cn}
\affiliation{Institute of High Energy Physics, Chinese Academy of Sciences, Beijing 100049, China}
\affiliation{Center for High Energy Physics, Peking University, Beijing 100871, China}

\author{Yu-Mei Wu}
\affiliation{Center for Gravitation and Cosmology, College of Physical Science
and Technology, Yangzhou University, Yangzhou 225009, China}

\author{Xiao Xue}
\affiliation{Institut de Física d’Altes Energies (IFAE), The Barcelona Institute of Science and Technology, Campus UAB, 08193 Bellaterra (Barcelona), Spain}

\author{Xingjiang Zhu}
\affiliation{Department of Physics,  Faculty of Arts and Sciences, Beijing Normal University, Zhuhai 519087, China}
\affiliation{ Institute for Frontier in Astronomy and Astrophysics, Beijing Normal University, Beijing 102206, China}

\begin{abstract}

Ultralight axion-like dark matter (ALDM) is a leading candidate in the dark matter realm, characterized by its prominent wave properties on astronomical scales. Pulsar timing arrays (PTAs) and Pulsar polarization arrays (PPAs) aim to detect this dark matter through timing and polarization measurements, respectively, of pulsars. The PTA relies on gravitational effects, as the ALDM halo perturbs the spacetime metric within the Milky Way, while the PPA detects nongravitational effects, namely cosmological birefringence induced by the ALDM Chern-Simons coupling with photons. These two methods complement each other, synergistically enhancing the pulsar array's capability to identify the ALDM signals in the data. In this article, we provide a foundational development of this synergy. We begin by revisiting previously derived two-point correlation functions for both PTA and PPA, and extend the analysis to include the leading‑order correlation between timing and polarization signals, which is encoded as a three‑point function. We then explore the constructions of likelihood functions for PTA and combined PTA-PPA analyses within a Bayesian framework, aimed at detecting the characteristic correlations of ALDM signals. We emphasize the non-Gaussianity of the ALDM timing signals, which arises from their nonlinear dependence on the field, in contrast to the Gaussian nature of its polarization signals. To address the complexities introduced, we approach this investigation in two ways: one involves a Gaussian approximation with proper justifications, while the other derives the formalism from the generic Gaussian characteristics of the ALDM field. Particularly, for the combined PTA-PPA case, a clear connection between the three-point correlation function and the likelihood is established. We anticipate that these efforts will lead to further developments in PTA and PTA-PPA analysis methods, better accounting for the influence of non-Gaussianity.

\end{abstract}

\maketitle
\newpage

\section{Introduction}


Dark matter (DM) is one of the biggest puzzles in fundamental science. 
Cosmological surveys reveal that DM comprises about 27\% of the Universe, compared to 5\% for ordinary baryonic matter, yet its nature remains elusive.
Axion-like dark matter (ALDM) is a leading DM candidate. Originally, the axion was introduced to resolve the strong charge-parity problem in quantum chromodynamics, but many theories in particle physics predict the existence of axion-like particles. If these bosonic particles have a mass $m_a \lesssim 30$\,\text{eV}, they are often referred to as ``wave DM'' (see~\cite{Hui:2021tkt} for a review). In this case, their de Broglie wavelength far exceeds their average spatial separation in the Milky Way (MW). 
This results in occupation numbers much greater than one within a de Broglie-scale volume, forming a coherent state or ``classical'' field.
Ultralight ALDM, with $m_a \lesssim 10^{-18}\,$eV~\cite{Marsh:2015xka}, stands out due to its prominent wave nature on astronomical scales. In particular, the variant with a mass $\sim 10^{-22}\,$eV,
commonly referred to as ``fuzzy DM''~\cite{1985MNRAS.215..575K, Hu:2000ke,hui2017ultralight}, has been proposed as a potential solution to small-scale structure problems in astronomy~\cite{weinberg2015cold,Hu:2000ke,hui2017ultralight}. The constraints on this DM scenario may arise from the observations of Lyman-$\alpha$ forest and dwarf galaxies (for a review, see~\cite{Ferreira:2020fam}). As these constraints could be subject to various systematic uncertainties~\cite{Zhang:2017chj,Ferreira:2020fam,Hayashi:2021xxu,Dalal:2022rmp,Zimmermann:2024xvd,Teodori:2025rul, Benito:2025xuh}, it is highly valuable to develop independent probes. Among these efforts, pulsar timing array (PTA)~\cite{hellings1983upper} and pulsar polarization array (PPA)~\cite{Liu:2021zlt} are especially promising.  

The PTA was originally proposed as a galactic-scale detector for nanohertz gravitational waves and stochastic gravitational wave backgrounds (SGWBs)~\cite{Sazhin:1978myk,detweiler1979pulsar,hellings1983upper,foster1990constructing}. In astronomy, millisecond pulsars (MSPs) are known for their long-term rotational stability, whose Hadamard or Allan variance can reach a level of $10^{-15}$, comparable to the current atomic clock standards (see, e.g.,~\cite{10.1093/mnras/stae331}). The SGWB reveals itself through Hellings-Downs correlations in pulsar pulse arrival times~\cite{hellings1983upper}, which thus can be detected by timing a group of MSPs and cross-correlating their  residuals~\cite{hellings1983upper}. 
Khmelnitsky and Rubakov noted that ultralight ALDM, with its gradient energy component accounting for $\sim 10^{-6}$ of the halo energy density, can perturb the MW metric with an oscillating pattern through minimal gravitational coupling with baryonic matter~\cite{Khmelnitsky:2013lxt}. This enabled searches for the ultralight ALDM based on identifying such patterns in PTA timing residuals~\cite{Khmelnitsky:2013lxt,DeMartino:2017qsa}. So far, several leading PTA collaborations including the Parkes PTA (PPTA)~\cite{Porayko:2018sfa}, NANOGrav~\cite{NANOGrav:2023hvm} and the European PTA (EPTA)~\cite{EuropeanPulsarTimingArray:2023egv} have delivered their first results of constraining the local energy density of ALDM in the Galactic halo, using their cutting-edge timing data. Recent searches for ALDM in PTA data also consider nonminimal coupling of ALDM to the Standard Model~\cite{Smarra:2024kvv,Kaplan:2022lmz}.

As highlighted in~\cite{Luu:2023rgg}, these analyses did  not fully incorporate the intricate structure of pulsar cross-correlation for the ALDM-induced timing signals. This information is expected to play a crucial role in recognizing the nature of any anomalous signals in the PTA detection of the ultralight ALDM, as it does for the SGWB detection. The wave characteristics of the ultralight ALDM in this context are encoded as correlation functions of timing residuals, a counterpart of the Hellings-Downs curve for SGWBs. To address this matter, some authors of this paper have derived the two-point correlation function of ALDM-induced timing residuals in~\cite{Luu:2023rgg}~\footnote{The two-point correlation function of ALDM-induced timing residuals was also calculated in~\cite{Boddy:2025oxn} recently. However, as  its authors agreed (private communications), the primary difference between~\cite{Boddy:2025oxn} and~\cite{Luu:2023rgg} in this calculation is that~\cite{Boddy:2025oxn} used the standard halo model (SHM) for the ALDM velocity distribution, while~\cite{Luu:2023rgg} assumed a delta function of speed. Notably, besides the calculation,~\cite{Luu:2023rgg}  also integrated this correlation function into its analysis framework and successfully conducted the first PTA analysis to search for the ALDM-induced correlations using real data in the same paper.}, and implemented it for the first time in the PTA analysis using the open $\gamma$-ray data of Fermi Large Area Telescope (for the application of $\gamma$-ray PTA for detecting the SGWBs, see~\cite{Fermi-LAT:2022wah}), under the assumption of a multivariate Gaussian distribution~\cite{Luu:2023rgg}. Note, another $\gamma$-ray PTA analysis of detecting the ultralight wave DM has also been performed roughly at the same time in~\cite{Xia:2023hov} which, however, ignored the pulsar cross-correlation 
as well.

The key role of pulsar cross-correlation in identifying the ultralight ALDM signals in pulsar data has been recognized even earlier for the PPA~\cite{Liu:2021zlt}. 
Timing and polarization are two essential features of pulsar pulses. To ensure high-precision measurement of the pulse time of arrival, astronomers often calibrate pulsar observations using polarization information. Thus, it was suggested in~\cite{Liu:2021zlt} to establish the PPA as a novel astronomical tool by cross-correlating polarization data acquired in the PTA programs. If astrophysics influences the light polarization of pulsars as a common signal correlated across galactic scales, it could be effectively investigated using the PPA. The ALDM field or halo within the MW, acting as a parity-odd background, can spontaneously break parity. This leads to a difference in the dispersion relation between left- and right-circular polarization modes of light due to its Chern-Simons coupling. Consequently, as linearly polarized pulsar light travels through the ALDM field, its position angle (PA) can rotate~\cite{Liu:2019brz}. This effect is generally known as ``cosmological birefringence (CB)''~\cite{Carroll:1989vb,Carroll:1991zs}. The wave nature of the ultralight ALDM predicts that the induced PA residuals are modulated as a common signal on astronomical scales~\cite{Liu:2021zlt}. The PPA, by cross-correlating the polarization data from different pulsars within the array thus can be highly capable of recognizing the ultralight ALDM CB signal. Following this proposal, the first PPA analysis of detecting the ultralight ALDM was performed in~\cite{Xue:2024zjq}, using the polarization data of 22 MSPs from the third data release of PPTA program. The  limits of the Chern-Simons coupling derived from this Bayesian analysis, where pulsar cross-correlation has been implemented, are found to be superior to the existing ones for the mass range of fuzzy DM. At the same time, the EPTA conducted an analysis of PPA data without modeling interpulsar correlations of ALDM signals~\cite{EPTA:2024gxu}.

Notably, the PTA infers gravitational effects by examining perturbations to the Galactic metric caused by ultralight ALDM, while the PPA addresses nongravitational effects, specifically the CB resulting from the ALDM's Chern-Simons coupling. Together, PTA and PPA offer complementary approaches for probing ultralight ALDM, and correlating timing and polarization data is expected to enhance the capabilities of pulsar arrays in distinguishing signals from noise. To achieve this, it is crucial to identify the characteristic correlation patterns of the ALDM timing and polarization signals, and to incorporate this information into a combined PTA-PPA Bayesian analysis framework. 

Nonetheless, developing Bayesian analysis framework requests a good understanding of statistical properties for the ALDM signals. One key issue is  non-Gaussianity of the timing signal, a characteristic that has been largely overlooked in previous studies. Unlike the ALDM-induced PA residuals, which depend linearly on the ALDM field, the ALDM-induced timing residuals rely on it quadratically, making it inherently non-Gaussian. 
Due to its non-Gaussian nature, deriving the exact likelihood function for Bayesian analysis of the entire array is challenging.
Nonetheless, the unique quadratic dependence on the field allows exploration of non-Gaussian distributions with certain approximations. For the PTA analysis, we show that the Gaussian approximation adopted in~\cite{Luu:2023rgg} could be a suitable starting point for incorporating the signal's rich correlation structure. For the combined PTA-PPA analysis, while a full likelihood is yet to be developed, we demonstrate how the leading-order three-point cross-correlation function can naturally emerge in  some minimal cases of the same analysis framework.

This paper is structured as follows. In Sec.~\ref{sec:correlation}, we revisit the characteristic correlations of the ALDM timing and polarization signals, respectively,  and then derive their cross-correlation functions. In Sec.~\ref{sec:PTAanalysis}, we examine the statistical properties of the ALDM timing signal and establish the PTA Bayesian analysis framework, validating the construction of the Gaussian likelihood function. We then expand our exploration to develop the PTA-PPA analysis framework, for proof of concept. Our findings are summarized in Sec.~\ref{sec:summary}. Additional details on statistical properties of the ALDM timing signal, as well as the derivations of formulas involving elementary Gaussian variables, are provided in Appendixes~\ref{app:1} and~\ref{app:2}, respectively.

\section{Correlations of ALDM Signals}
\label{sec:correlation}

The DM halos originate from primordial density fluctuations. However, given the evolution they have undergone -- through processes such as virialization, fragmentation, randomization and thermalization -- the ALDM halos as a classical field can be locally modeled as a random superposition of a large number of particle plane waves~\cite{Derevianko:2016vpm, Foster:2017hbq, Foster:2020fln,Cheong:2024ose}~\footnote{ One can replace particle plane waves in the superposition with their energy eigenstates, where gravitational potential within  galaxies has been fully considered, to achieve a more accurate modeling for the ALDM halos.}
\begin{eqnarray}\label{eq:axionf1}
a(\mathbf{x},t) \approx \frac{\sqrt{\rho(\mathbf{x})}}{m_a}   \sum_{\mathbf{v}\in\Omega}(\Delta v)^{3/2}\alpha_{\mathbf{v}}\sqrt{f (\mathbf{v})} \cos[\omega t-\mathbf{k}\cdot\mathbf{x}+\phi_{\mathbf{v}}]\,. 
\end{eqnarray}
Here $\mathbf{v}\in\Omega$ denotes lattice sites in phase space, and $\Delta v$ is their spacing. $\omega= m_a/\sqrt{1-v^2}$ and $\mathbf{k}= m_a \mathbf{v}/\sqrt{1-v^2}$ are wave angular velocity and vector. $f (\mathbf{v})$, assumed to be universal in space, describes the ALDM velocity distribution. For illustration, below we employ the  SHM~\cite{Drukier:1986tm, Evans:2018bqy}:  
\begin{equation}   \label{eq: SHM}
    f\left( \mathbf{v} \right) =\frac{1}{\pi ^{3/2}v_{0}^{3}}\exp \left[ -\frac{\left( \mathbf{v}+\mathbf{v}_{\odot} \right) ^2}{v_{0}^{2}} \right] \, ,
\end{equation}
where $v_0\approx 220\,\mathrm{km/s}$ is Galactic virial velocity and $\mathbf{v}_\odot\approx\{11, 232, 7\}\,$km/s is the Sun's velocity relative to the halo. Both $v_0$ and $|\mathbf{v}_\odot|$ are $\sim 10^{-3}$ in natural units, indicating that the ALDM is highly nonrelativistic. 
$\alpha_{\bf{v}}$ and $\phi_{\bf{v}}$ are random amplitude and phase parameters of the ALDM, arising from its stochastic nature and sampled at each lattice site from the Rayleigh distribution with a scale parameter $\sigma=1$, and from a uniform distribution, respectively. With these stochastic parameters, we can define a set of independent Gaussian basis (see discussions in Appendix~\ref{app:1}): $\{ \alpha_{\mathbf{v}} \cos \left( \phi _{\mathbf{v}} \right), \alpha_{\mathbf{v}} \sin \left(\phi _{\mathbf{v}} \right)| \mathbf{v}\in\Omega \}$. The ALDM field $a(\mathbf{x},t)$ can be linearly decomposed in this basis and is thus random Gaussian.\footnote{This discussion also implies that any linear combination of the ALDM field profiles should remain random Gaussian, regardless of whether these profiles are correlated or not. } 
Its statistical properties are fully characterized by its ensemble mean and covariance matrix then. The ALDM field in the MW as a consequence can be viewed as a specific realization following this Gaussian distribution, where $\rho(\bf{x})$ represents an ensemble average of the ALDM energy density at position $\mathbf{x}$.

The superposition of particle waves in phase space results in the stochastic time and space dependence of the ALDM field. At two points separated by $\tau$ in time and $\bf{d}$ in space, the ALDM field loses its coherence when $\tau\gg \tau_c$ or $|{\bf d}|\gg l_c$. Here $\tau_c \sim 1/(m_av_0^2)$ and $l_c\sim 1/(m_a v_0)$ denote coherent time and coherent length (i.e. de Broglie wavelength) of this field, respectively. In the nonrelativistic limit, the phase factor in Eq.~(\ref{eq:axionf1}) can be expanded as: $\omega t - \mathbf{k}\cdot\mathbf{x} = m_a(t -\mathbf{v}\cdot\mathbf{x} +\frac{1}{2}v^2t + \mathcal O(v^3))$. The observation time span $T_{\rm obs}$ is much shorter than $\tau_c$ for the ultralight ALDM. The $\frac{1}{2}m_a v^2 t$ and higher-order terms from the $\omega t$ expansion become negligible, and the ALDM temporal profile is thus well-described by a coherent time evolution of $\cos(m_a t+...)$. However, the pulsar distance, either to another pulsar or to the Earth, could be comparable to or even smaller than the coherence length $l_c$. The term of $m_a \mathbf{v}\cdot\mathbf{x}$ in the $\mathbf{k}\cdot\mathbf{x}$ expansion thus should be retained to capture the spatial information of the ALDM waves. Then with the higher-order spatial terms also neglected,\footnote{In the mass regime $m_a > 10^{-18}\,$eV, the ALDM can still be explored through its wave properties by examining the higher-order temporal and spatial terms from the phase expansion in Eq.~(\ref{eq:axionf1}). For example, higher-order temporal terms can lead to low-frequency fluctuations, as discussed in \cite{Kim:2023pkx,Kim:2023kyy}, while higher-order spatial terms might generate additional pulsar cross-correlations on larger astronomical scales. In the current context, however, these terms do not have a noticeable impact on the detection of ultralight ALDM with $m_a < 10^{-18}\,$eV.} 
we have~\cite{Foster:2017hbq, Liu:2021zlt}
\begin{eqnarray}\label{eq:axionf2}
a(\mathbf{x},t) &\approx& \frac{\sqrt{\rho \left( \mathbf{x} \right)}}{m_a}\sum_{\mathbf{v}\in \Omega}
(\Delta v ) ^{3/2}\alpha_{\mathbf{v}}\sqrt{f(\mathbf{v})} \cos\mathrm{[}m_a(t-\mathbf{v}\cdot \mathbf{x})+\phi _{\mathbf{v}}]\nonumber\\
&\equiv& \frac{\sqrt{\rho \left( \mathbf{x} \right)}}{m_a}\sum_{\mathbf{v}\in \Omega}\mC_{\mathbf{v}}\cos[\vartheta_{\mathbf{v}}(\mathbf x,t)]\, ,
\end{eqnarray}
where $\mC_{\mathbf{v}}\equiv (\Delta v ) ^{3/2}\alpha_{\mathbf{v}}\sqrt{f(\mathbf{v})}$ is a shorthand notation for the coefficient and $\vartheta_{\mathbf{v}}(\mathbf x,t)\equiv m_a(t-\mathbf{v}\cdot \mathbf{x})+\phi_{\mathbf{v}}$ is a phase parameter.
Determining the value of $\rho(\bf{x})$ is subtle because the MW only warrantees one realization of $\{\alpha_{\bf{v}}, \phi_{\bf{v}}\}$. However, we notice that its variance over the ensemble is not excessively large, by sampling the random parameters $\{\alpha_{\bf{v}}, \phi_{\bf{v}}\}$. So at leading order, we can approximate it with the measured DM energy density in the MW. Its value is usually extracted out from the rotation curve within the MW inner region which has a diameter $\sim 20\,$kpc~\cite{Nesti:2013uwa}. To ensure that this measurement covers at least multiple coherent volumes of the ALDM field, thereby unbiased by density fluctuations caused by interference, we consider ALDM with $m_a \gtrsim 10^{-23.5}\,$eV, where $l_c \lesssim 1.9\,$kpc. 
This defines the relevant mass range for this research as $10^{-23.5}\, {\rm eV} \lesssim m_a \lesssim 10^{-18}\,$eV.

Next we will review the two-point correlation functions of the ALDM signals which were first derived in~\cite{Liu:2021zlt} for PA residuals and in~\cite{Luu:2023rgg} for timing residuals, and then investigate the correlations between the ALDM timing and polarization signals.

\subsection{Polarization signal and two-point correlation functions}

The ALDM can interact with pulsar light through the Chern-Simons term $\sim \frac{1}{2}g_{a\gamma\gamma}\,a\,F_{\mu\nu}\tilde{F}^{\mu\nu}\,$, where $F_{\mu\nu}$ is the electromagnetic field strength, $\tilde{F}_{\mu\nu}$ is its Hodge dual, and $g_{a\gamma\gamma}$ is the Chern-Simons coupling. Due to the topological nature of $F_{\mu\nu}\tilde{F}^{\mu\nu}$, and thus CB, the ALDM-induced PA residual depends only on the field profile at the endpoints of the light path. For a pulse emitted by the pulsar at $(\mathbf{x}_p,t_p)$ and received on the Earth at $(\mathbf{x}_e,t_e)$, this PA residual is given by
\begin{eqnarray}\label{eq:DPA0} \dPA^a=g_{a\gamma\gamma}\left[a(\mathbf{x}_p,t_p)-a(\mathbf{x}_e,t_e)\right]\,.
\end{eqnarray}
namely a ``pulsar'' term and an ``Earth'' term together.  
For a pulsar array with $\mathcal{N}$ pulsars, one can construct a vector of the ALDM-induced PA residuals:  
\begin{equation}\label{eq:DPAvec}
    \bDPA^a=\left(\dPA^a_{1,1},\ldots, \dPA^a_{1,N_1},\ldots,  \dPA^a_{p,1}, \ldots, \dPA^a_{p,N_p}, \ldots,  \dPA^a_{\mathcal{N},1},\ldots, \dPA^a_{\mathcal{N},N_{\mathcal{N}}}\right)^T\,,
\end{equation}
where $n$ and $p$ denote the $n$-th epoch of the $p$-th pulsar, and $N_p$ represents the number of observation epochs for the $p$-th pulsar. Specifically, with the ALDM field in Eq.~(\ref{eq:axionf2}), $\dPA^a_{p,n}$ is given by 
\begin{eqnarray}\label{eq:DPA1}
\dPA_{p,n}^a&=&\frac{g_{a\gamma\gamma}}{m_a}\sum_{\mathbf{v}\in \Omega}\mC_{\mathbf{v}}\Big\{\sqrt{\rho_p}\cos[m_a (t_{p,n}-L_p-\mathbf{v}\cdot\mathbf{x}_p)+\phi_{\mathbf{v}}]-\sqrt{\rho_e}\cos[m_a t_{p,n}+\phi_{\mathbf{v}}]\Big\} \, ,
\end{eqnarray}
where $\rho_p=\rho(\mathbf{x}_p)$ and $\rho_e=\rho(\mathbf{x}_e)$ are the halo densities around the $p$-th pulsar and near the Earth, respectively. Here, $t_{p,n}$ denotes the pulse arrival time on Earth at the $n$-th epoch for this pulsar, and $L_p=|\mathbf{x}_p-\mathbf{x}_e|$ is the distance of the pulsar to Earth (i.e. $t_p$ for the $n$-th epoch in Eq.~(\ref{eq:DPA0}) is $t_{p,n}-L_p$). As $\dPA^a_{p,n}$ is a linear combination of the ALDM field profiles, it respects Gaussian statistics. Accordingly, $\bDPA^a$ follows a multivariate Gaussian distribution with a zero mean.  

For the convenience of later discussions, we express the PA residual in a compact form: 
\begin{eqnarray}\label{eq:DPA1c}
\dPA_{p,n}^a =-\frac{g_{a\gamma\gamma}}{m_a}\sum_{i=0,1}(-1)^i \sqrt{\rho(\mathbf{x}_p^{(i)})}X_{p,n}^{(i)}\,,
\end{eqnarray}
where $\mathbf{x}_{p}^{(0)}=\mathbf{x}_e=0$, $t_{p,n}^{(0)}=t_{p,n}$ and $\mathbf{x}_{p}^{(1)}=\mathbf{x}_p$, $t_{p,n}^{(1)}=t_{p,n}-L_p$. $X_{p,n}^{(i)}$ is a Gaussian variable related to the ALDM field, i.e. 
\begin{eqnarray}\label{eq:Xpn}
    X_{p,n}^{(i)}
    =&{}\sum_{\mathbf{v}\in \Omega}\mC_{\mathbf{v}}\cos\Big[\vartheta_{\mathbf{v}}(\mathbf{x}_p^{(i)},t_{p,n}^{(i)})\Big]\,, 
\end{eqnarray}
with the amplitude and phase parameters $\mC_{\mathbf{v}}$ and $\vartheta_{\mathbf{v}}(\mathbf x,t)$ defined in Eq.~(\ref{eq:axionf2}). Its statistical properties are then inherited from ALDM. Since $\phi_{\mathbf{v}}$ follows a uniform distribution, the ensemble mean $\langle X_{p,n}^{(i)}\rangle$ is zero.
The vector $\bX^{(i)}=(X^{(i)}_{1,1},..., X^{(i)}_{p,n},...,X^{(i)}_{\mathcal{N},N_\mathcal{N}})^T$ then follows a multivariate Gaussian distribution with zero mean.  Its covariance matrix, $\bC_X^{(ij)}=\langle \bX^{(i)} (\bX^{(j)})^T \rangle$, is symmetric with respect to $i$ and $j$. 
Using the velocity distribution given in Eq.~(\ref{eq: SHM}), the entries of the covariance matrix can be derived as (see Appendix~\ref{app:2} for details)
\begin{eqnarray}\label{eq:bCXij}
  (\bC_X^{(ij)})_{pn,qm}
  = e^{-\frac{1}{4}(y_{pq}^{ij})^2} \cos \Big[ m_a ( t_{p,n}^{\left( i \right)}-t_{q,m}^{\left( j \right)}) +m_a\mathbf{v}_{\odot}\cdot \mathbf{x}_{pq}^{\left( ij \right)} \Big]\,, 
\end{eqnarray}
where $\mathbf{x}_{pq}^{(ij)}\equiv \mathbf{x}_p^{(i)}-\mathbf{x}_q^{(j)}$ and  $y^{ij}_{pq}\equiv |\mathbf{x}^{(ij)}_{pq}|/l_c$.

In view of its Gaussian nature, the statistical information of $\bDPA^a$ is fully encoded in the covariance matrix 
\begin{eqnarray}\label{eq:CPAa}
    \bC_{\rm PA}^{a}=\langle \bDPA^a (\bDPA^a)^T\rangle
    =\frac{g_{a\gamma\gamma}^2}{m_a^2}\sum_{i,j}(-1)^{i+j}\sqrt{\rho(\mathbf{x}_p^{(i)})\rho(\mathbf{x}_q^{(j)})}\, \bC_{X}^{(ij)} \, ,  
\end{eqnarray}
with its entries $(\bC_{\rm PA}^{a})_{pn,qm}$ defined by the two-point correlation functions:   
\begin{equation}\label{eq:DPAcorr1}
\langle \dPA_{p,n}^{a} \dPA_{q,m}^{a}\rangle 
    = \frac{g_{a\gamma\gamma}^2}{m_a^2}\sum_{i,j}(-1)^{i+j}\sqrt{\rho(\mathbf{x}_p^{(i)})\rho(\mathbf{x}_q^{(j)})}e^{-\frac{1}{4}(y_{pq}^{ij})^2} \cos \Big[ m_a ( t_{p,n}^{\left( i \right)}-t_{q,m}^{\left( j \right)}) +m_a\mathbf{v}_{\odot}\cdot \mathbf{x}_{pq}^{\left( ij \right)} \Big]\,. 
\end{equation}
Here $i$ and  $j$ together run over four possible correlation modes for the signal: Earth-Earth, Earth-Pulsar, Pulsar-Earth and Pulsar-Pulsar. They can be expressed in a more explicit form:  
\begin{eqnarray}\label{eq:DPAcorr2}
    \langle  \dPA_{p,n}^{a} \dPA_{q,m}^{a}\rangle 
    &=& \frac{g_{a\gamma\gamma}^2}{m_a^2}\bigg\{\rho_e\cos\Big[m_a \Delta t_{pn,qm} \Big]  \nonumber \\ 
&& + \sqrt{\rho_p\rho_q}\cos\Big[m_a(\Delta t_{pn,qm}- L_{pq}+\mathbf{v}_{\odot}\cdot \mathbf{x}_{pq}) \Big]e^{-\frac{1}{4} y_{pq}^2}\nonumber\\
    &&-\sqrt{\rho_e\rho_p}\cos\Big[m_a(\Delta t_{pn,qm}-L_p+\mathbf{v}_{\odot}\cdot \mathbf{x}_{pe}) \Big]e^{-\frac{1}{4}y_{ep}^2}\nonumber\\
    &&-\sqrt{\rho_e\rho_q}\cos\Big[m_a(\Delta t_{pn,qm}+L_q+\mathbf{v}_{\odot}\cdot \mathbf{x}_{eq}) \Big]e^{-\frac{1}{4}y_{eq}^2}
    \bigg\}\,,
\end{eqnarray}
where $\mathbf{x}_{ij}=\mathbf{x}_i-\mathbf{x}_j$, $y_{ij}=|\mathbf{x}_{ij}|/l_c$, $\Delta t_{pn,qm}= t_{p,n}-t_{q,m}$ and $L_{pq}= L_p-L_q$. 

In these two-point correlation functions, trigonometric factors describe temporal correlations of the ALDM signals, with one additional term in phase introduced to account for the solar velocity relative to the halo $\mathbf{v}_{\odot}$.
The spatial correlation of the ALDM signals are captured by exponential factors $e^{-\frac{1}{4}(y_{pq}^{ij})^2}$ (or $e^{-\frac{1}{4}(y_{ij})^2}$ for Eq.~(\ref{eq:DPAcorr2})), which becomes important for 
$|\mathbf{x}^{(ij)}_{pq}| \lesssim l_c$ (or $|\mathbf{x}_{ij}| \lesssim l_c$ for Eq.~(\ref{eq:DPAcorr2})). This effect is encoded as a sinc function in Ref.~\cite{Liu:2021zlt,Xue:2024zjq}, where the DM speed distribution $f(v)$ is modeled with a delta function. While these two functions differ in form, they predict similar features regarding signal spatial correlations. In the large $m_a$ regime, where the ALDM de Broglie wavelength $l_c$ becomes smaller than the length scale of pulsar array, resulting in suppressed spatial correlation, the temporal correlation in the Earth-Earth term can still play an important role in identifying the ALDM signals. This holds true until the signal oscillation period, determined by $1/m_a$, becomes shorter than the interval between consecutive observation epochs.

\subsection{Timing signal and two-point correlation functions} \label{subsec:PTAcf}

As first shown in \cite{Khmelnitsky:2013lxt}, the ALDM field perturbs gravitational potential within a galaxy. In the Newtonian gauge, this effect can be represented as $h_{ij} = 2\Psi\delta_{ij}$, where $\Psi$ is a scalar potential. The ALDM dynamical pressure $p(\mathbf{x},t)$ then introduces an oscillating component $\Psi_c (\mathbf{x},t)$ in the scalar potential, described by 
\begin{eqnarray}\label{eq:PsicEQ}
    -6\ddot\Psi_c (\mathbf{x},t)\approx 24\pi G p(\mathbf{x},t)\approx 12\pi G \left[\dot{a}^2(\mathbf{x},t)-m_a^2a^2(\mathbf{x},t)\right]\, ,
\end{eqnarray}
where the spatial-derivative terms have been neglected due to nonrelativistic suppression. By substituting the ALDM field from Eq.~(\ref{eq:axionf2}) into Eq.~(\ref{eq:PsicEQ}), we obtain 
\begin{eqnarray} \label{eq:Psi}
\Psi_c (\mathbf{x},t)&\approx&\frac{\pi G}{2m_a^2}\left[m_a^2a^2(\mathbf{x},t)-\dot{a}^2(\mathbf{x},t)\right]\,.
\end{eqnarray}
This perturbation can further induce timing residuals for pulsar pulses: 
\begin{eqnarray}\label{eq:res}
\dt^a(t)&=&-\int_{t_0}^t\frac{\nu(t')-\nu_0}{\nu_0}dt' \approx  -\int_{t_0}^{t} \left[\Psi_c(\mathbf{x}_p,t_p')- \Psi_c(\mathbf{x}_e,t') \right]\,d t'\nonumber\\
&\approx& -\frac{\pi G}{2m_a^2}\Big[\dot{a}(\mathbf{x}_p,t-L_p)a(\mathbf{x}_p,t-L_p)-\dot{a}(\mathbf{x}_e,t)a(\mathbf{x}_e,t)\Big]+{\rm const}  \, .
\end{eqnarray}
Here, $t_0$ is a reference time, $t_p'\approx t'-L_p$ and $t'$ are pulse emission and arrival moments, and $\nu_0$ and $\nu(t')$  represent pulse intrinsic and apparent frequencies. The pulsar timing residual measures the relative frequency shift cumulated along its light path. The choice of $t_0$ may introduce a constant offset to the timing residual time series. Considering that the usual PTA analysis will marginalize such an offset as an unknown deterministic noise, we will simply remove it in the following discussion. The timing signals  manifest as a difference between one ``pulsar'' term and one ``Earth'' term also, which are related to the ALDM perturbations at the end points of the light path respectively.  

Then, we can construct a vector of the ALDM-induced timing residuals for a pulsar array:
\begin{eqnarray}\label{eq:Dtvec}
    \bDt^a=(\dt^a_{1,1},...,\dt^a_{1,N_1},...,\dt^a_{p,1},...,\dt^a_{p,N_p},...,\dt^a_{\mathcal{N},1},...,\dt^a_{\mathcal{N},N_{\mathcal{N}}})^T\, ,
\end{eqnarray}
where $p$ and $n$ again denote the $n$-th epoch of the $p$-th pulsar. Specifically, $\dt^a_{p,n}$ is given by 
\begin{equation}\label{eq:dti1}
\begin{aligned}
\dt^a_{p,n}=-\frac{\pi G}{4m_{a}^{3}}&\left\{ \rho_p\sum_{\mathbf{v},\mathbf{v}^{'}\in \Omega}\mC_{\mathbf{v}}\mC_{\mathbf{v}'}\sin\mathrm{[}2m_a(t_n-L_p)-m_a(\mathbf{v}+\mathbf{v}^{'})\cdot \mathbf{x}_p+\phi _{\mathbf{v}}+\phi _{\mathbf{v}^{'}}]\right.\\
&\left.-\rho_e\sum_{\mathbf{v},\mathbf{v}^{'}\in \Omega}\mC_{\mathbf{v}}\mC_{\mathbf{v}'}\sin\mathrm{[}2m_at_n+\phi _{\mathbf{v}}+\phi _{\mathbf{v}^{'}}] \right\} \, .
\end{aligned}	
\end{equation}
Unlike its polarization signal, which has a linear dependence on the ALDM field profile, the ALDM-induced timing residual relies on the field profile quadratically, making it inherently non-Gaussian.

As in the case of PA residuals, it is useful to express the timing residual in a compact form:
\begin{eqnarray}\label{eq:DtXY}
\dt^a_{p,n} = \frac{\pi G}{2m_{a}^{3}}\sum_{i=0,1}{\left( -1 \right) ^i\rho( \mathbf{x}_{p}^{\left( i \right)}) X_{p,n}^{(i)} Y_{p,n}^{(i)}}\, ,
\end{eqnarray}
where $X_{p,n}^{(i)}$ is defined in Eq.~(\ref{eq:Xpn}) and $Y_{p,n}^{(i)}$ represents another class of Gaussian variables with zero mean, i.e. 
\begin{eqnarray}\label{eq:Ypn}
Y_{p,n}^{(i)}\equiv  \sum_{\mathbf{v}\in \Omega}\mC_{\mathbf{v}}\sin\Big[\vartheta_{\mathbf{v}}(\mathbf{x}_p^{(i)},t_{p,n}^{(i)})\Big]\,, 
\end{eqnarray}
with the same $\mC_{\mathbf{v}}$ and $\vartheta_{\mathbf{v}}(\mathbf x,t)$ defined in Eq.~(\ref{eq:axionf2}). The statistical properties of $\bDt^a$ are thus fully determined by the covariance matrices for the vectors $\bX^{(i)}$ and $\bY^{(i)}$, {\it i.e.},  $\bC_X^{(ij)}=\langle \bX^{(i)} (\bX^{(j)})^T \rangle= \langle \bY^{(i)} (\bY^{(j)})^T \rangle$ in Eq.~(\ref{eq:bCXij}), and $\bC_{XY}^{(ij)}=\langle \bX^{(i)} (\bY^{(j)})^T \rangle$ with the entries (see Appendix~\ref{app:2} for details)  
\begin{eqnarray}\label{eq:bCXYij}
(\bC_{XY}^{(ij)})_{pn,qm}
=-e^{-\frac{1}{4}(y_{pq}^{ij})^2} \sin \Big[ m_a ( t_{p,n}^{\left( i \right)}-t_{q,m}^{\left( j \right)}) +m_a\mathbf{v}_{\odot}\cdot \mathbf{x}_{pq}^{\left( ij \right)} \Big]\,.
\end{eqnarray}
Note that $\bC_{XY}^{(ij)}$ is antisymmetric, indicating $\langle X_{p,n}^{(i)}Y_{p,n}^{(i)} \rangle=0$.  Thus, the ensemble mean of $\bDt^a$ is zero. 

The statistical information at leading order emerges as two-point correlation functions, which are given by
\begin{eqnarray}\label{eq:Dtcorr}     
\left< \dt_{p,n}^{a}\dt_{q,m}^{a} \right>
&=& \frac{\pi ^2G^2}{4m_{a}^{6}}\sum_{i,j}{\left( -1 \right) ^{i+j}}\rho( \mathbf{x}_{p}^{\left( i \right)}) \rho(\mathbf{x}_{q}^{\left( j \right)})
\left\langle X_{p,n}^{(i)} Y_{p,n}^{(i)} X_{q,m}^{(j)} Y_{q,m}^{(j)} \right\rangle \\
&=&\frac{\pi ^2G^2}{4m_{a}^{6}}\sum_{i,j}{\left( -1 \right) ^{i+j}\rho ( \mathbf{x}_{p}^{\left( i \right)}) \rho(\mathbf{x}_{q}^{\left( j \right)})}e^{-\frac{1}{2}\left( y_{pq}^{ij} \right) ^2}\cos \Big[ 2m_a\left( t_{p,n}^{(i)}-t_{q,m}^{(j)} \right) +2m_a\mathbf{v}_{\odot}\cdot  \mathbf{x}_{pq}^{(ij)} \Big] \,, \nonumber 
\end{eqnarray}
where we have applied Eqs.~(\ref{eq:bCXij}) and (\ref{eq:bCXYij}).
Consequently, the covariance matrix for the timing signal vector can be expressed as
\begin{eqnarray}\label{eq:Ctafull}
  \bC_{t}^{a}=\langle \bDt^a (\bDt^a)^T\rangle
    =\frac{\pi^2 G^2}{4m_a^6}\sum_{i,j}(-1)^{i+j}\rho ( \mathbf{x}_{p}^{\left( i \right)}) \rho(\mathbf{x}_{q}^{\left( j \right)})\left[\bC_{X}^{(ij)}\odot \bC_{X}^{(ij)}- \bC_{XY}^{(ij)}\odot\bC_{XY}^{(ij)}\right]\,,
\end{eqnarray}
where $\odot$ denotes Hadamard product, {\it i.e.},  $(A\odot B)_{mn}=A_{mn}B_{mn}$. Compared to the covariance matrix for the PA residuals in Eq.~(\ref{eq:CPAa}), $\bC_{t}^{a}$ exhibits a more complex structure,  highlighting the potential significance of incorporating the full correlation information in the PTA analysis. This two-point correlation function can be also expressed in a more explicit form 
\begin{eqnarray}\label{eq:Dtcorr2}
\left< \dt_{p,n}^{a}\dt_{q,m}^{a} \right>
&=&\frac{\pi ^2G^2}{4m_{a}^{6}}
\bigg\{\rho_e^2 \cos\Big[ 2m_a \Delta t_{pn,qm} \Big] \nonumber \\ 
&&+\rho_p\rho_q \cos\Big[2m_a(\Delta t_{pn,qm}-L_{pq}+\mathbf{v}_{\odot}\cdot  \mathbf{x}_{pq})\Big]e^{-\frac{1}{2}y_{pq}^2}\nonumber\\
&&-\rho_e\rho_p \cos\Big[2m_a(\Delta t_{pn,qm}-L_p+\mathbf{v}_{\odot}\cdot \mathbf{x}_{pe})\Big]e^{-\frac{1}{2}y_{ep}^2}\nonumber\\
&&-\rho_e\rho_q \cos\Big[2m_a(\Delta t_{pn,qm}+L_q-\mathbf{v}_{\odot}\cdot  \mathbf{x}_{qe})\Big]e^{-\frac{1}{2}y_{qe}^2}\bigg\} \,.
\end{eqnarray}
Similar to the polarization case, in these two-point correlation functions trigonometric factors explain temporal correlations of the ALDM signals, and exponential factors account for their spatial correlations. Because of the quadratic dependence of the timing residuals on the ALDM field, the characteristic scale is reduced by half for the temporal correlations, and in the $m_a$ regime where $l_c$ is shorter than the length scale of pulsar arrays, the suppression rate for spatial correlations is doubled.  

As the PPA method does~\cite{Liu:2021zlt,Xue:2024zjq}, the PTA's approach to searching for ALDM correlations differs significantly from its response to nanohertz SGWB~\cite{Luu:2023rgg}. In the latter case, the Hellings-Downs curve arises from the Earth-Earth correlation term and receives only subleading contributions from the pulsar-related terms. 
This is because the wavelength of GWs, which travel at the speed of light, is much shorter than both the distances from the pulsars to Earth and the distances between the pulsars themselves. In contrast, the ALDM de Broglie wavelength $l_c$ is enhanced by a factor $\sim 1/v_0\approx 10^{3}$ compared to $1/m_a$, due to its nonrelativistic nature. Consequently, all four correlation terms in the two-point correlation functions are accessible to the PTA for the mass regime of ``fuzzy DM'' (where $m_a$ falls into the nanohertz frequency band) and could play a key role for identifying its signals at leading order.

\subsection{Correlations between timing and polarization signals}
\label{subsec:ccorr}

The ALDM-induced PA residuals (see Eq.~(\ref{eq:DPA1})) and timing residuals (see Eq.~\eqref{eq:dti1}) are generically correlated, due to their common origin, with specific patterns in spacetime. In contrast, the PTA and PPA noises are mostly  uncorrelated or correlated but with different characteristic patterns. To fully utilize the data of pulsar array to investigate the ultralight ALDM, we can correlate its timing and polarization signals, thereby achieving a synergy of gravitational (PTA) and nongravitational (PPA) methods.   

The ALDM-induced PA and timing residuals have a zero ensemble mean, as discussed above. Their two-point correlation functions are also zero, {\it i.e.}, 
\begin{eqnarray}
    \langle \dt_{p,n}^{a} \dPA_{q,m}^{a}\rangle
    =
    -\,\frac{\pi G g_{a\gamma\gamma}}{4m_a^4} \sum_{i,j}(-1)^{i+j}\rho(\mathbf{x}_p^{(i)})\sqrt{\rho(\mathbf{x}_q^{(j)})}\left\langle X_{p,n}^{(i)} Y_{p,n}^{(i)} X_{q,m}^{(j)}\right\rangle=0\,,
\end{eqnarray}
since the ensemble mean of an odd number of Gaussian variables is zero. Therefore, the leading-order cross-correlation between the polarization and timing signals must involve two PA residuals and one timing residual, manifested as a three-point correlation function.  
By utilizing the velocity distribution provided in Eq.~(\ref{eq: SHM}), we derive the correlation function as  (see Appendix~\ref{app:2} for details)
\begin{eqnarray}\label{eq:DPADPADtcorr}
    \langle \dPA^a_{p,n} \dPA^a_{q,m} \dt^a_{r,l}\rangle
    &=&
    \frac{\pi G g^2_{a\gamma\gamma}}{2m_a^5} \sum_{i,j,k}(-1)^{i+j+k}\sqrt{\rho(\mathbf{x}_p^{(i)})\rho(\mathbf{x}_q^{(j)})}\rho(\mathbf{x}_r^{(k)})\left\langle X_{p,n}^{(i)}X_{q,m}^{(j)}X_{r,l}^{(k)}Y_{r,l}^{(k)}\right\rangle\nonumber\\
    &=&-\frac{\pi Gg_{a\gamma \gamma}^{2}}{2m_{a}^{5}}\sum_{i,j,k}{\left( -1 \right) ^{i+j+k}}\sqrt{\rho(\mathbf{x}_p^{(i)})\rho(\mathbf{x}_q^{(j)})}\rho(\mathbf{x}_r^{(k)})\, e^{-\frac{1}{4}\left( y_{pr}^{ik} \right) ^2} e^{-\frac{1}{4}\left( y_{qr}^{jk} \right) ^2 }\nonumber\\
    && \times \sin \Big[ m_a( t_{p,n}^{\left( i \right)}+t_{q,m}^{\left( j \right)}-2t_{r,l}^{\left( k \right)}) +m_a\mathbf{v}_{\odot}\cdot (\mathbf{x}_{pr}^{\left( ik \right)}+ \mathbf{x}_{qr}^{\left( jk \right)}) \Big]\,.
\end{eqnarray} 
Here, spatial correlations are described by a product of two exponential factors, namely $e^{-\frac{1}{4}(y_{pr}^{ik})^2}e^{-\frac{1}{4}(y_{qr}^{jk})^2}$, each representing a correlation between one PA residual and one Gaussian variable from the timing residual. Temporal correlations encoded in trigonometric factors also reflect this effect in their phase structure.   These features arise from the way polarization and timing signals are composed of the two sets of Gaussian variables in Eqs.~(\ref{eq:DPA1c}) and (\ref{eq:DtXY}). Incorporating the three-point functions into the Bayesian  analysis framework is a complex task. We will perform an exploratory study regarding this in the next section.

\section{Data Analysis Methodology}
\label{sec:PTAanalysis}

As discussed above, the timing residuals induced by the ALDM exhibit more complex statistical properties than PA residuals due to their nonlinear dependence on the field. Below, we will examine these statistical properties in detail and develop a PTA Bayesian analysis framework that properly incorporates the correlation features of the timing signals. The discussion will expand to include the combined PTA-PPA Bayesian analysis then.

\subsection{Statistical properties of ALDM timing signals}
\label{sec:Dtstat}

To examine the statistical properties of the ALDM-induced residuals, let us consider two limits: (1) $\rho_p \gg \rho_e$, where the pulsars are close to the Galactic center and the environmental DM density is expected to be dense, and (2) $\rho_p \approx \rho_e$, where the pulsars are not far from the Earth (with a distance $\lesssim \mathcal O(1)\,$kpc), as is the case with current PTA constructions, and the DM density is approximately uniform. 

\begin{figure}[h]
    \centering
    \includegraphics[height=5.0cm]{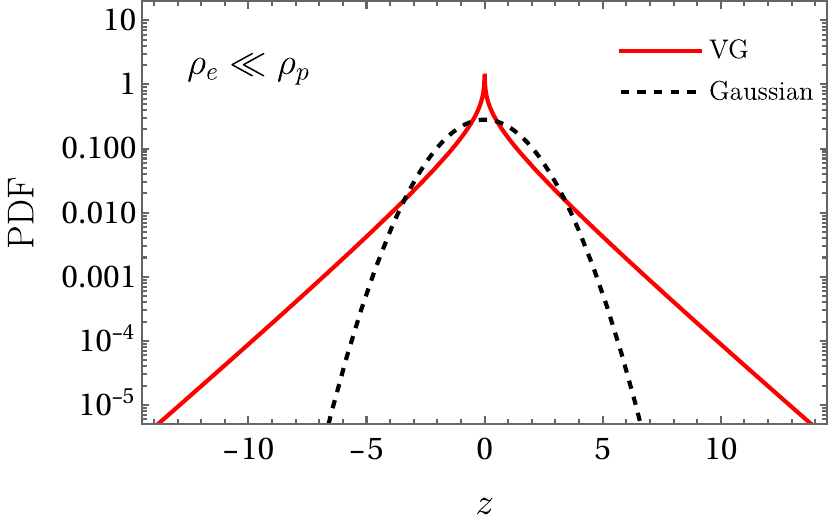}\;\;
    \includegraphics[height=5.0cm]{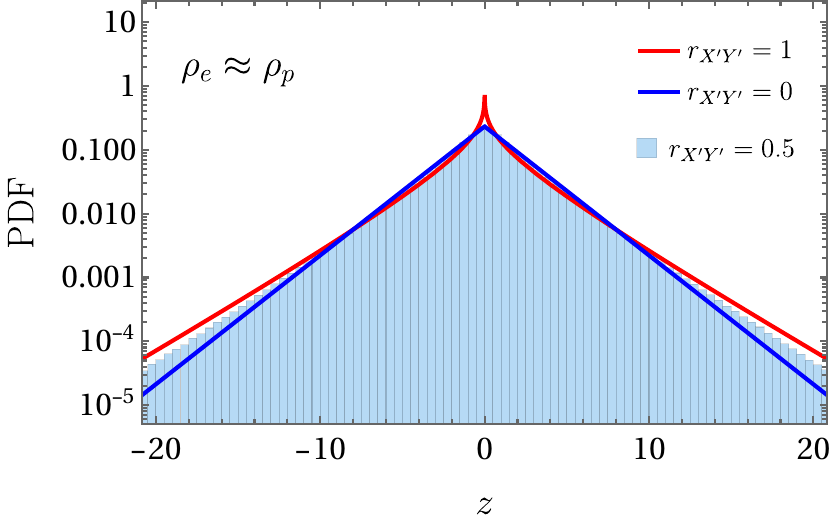}
    \captionsetup{justification=raggedright, singlelinecheck=false}
    \caption{Left: comparison of the VG distribution   in Eq.~(\ref{eq:VGdis}) for $\dt$ in the limit of $\rho_p\gg\rho_e$ ($\sigma^2=2$) with a Gaussian distribution of identical variance. Right: distribution of $\dt$ in the limit of $\rho_p\approx \rho_e$, for $r_{\tilde X \tilde Y} =$ 0, 0.5 and 1 ($\sigma_{\tilde X}=\sigma_{\tilde Y}=1$). We present analytical results in Eq.~(\ref{eq:fDtlimit}) for $r_{\tilde{X} \tilde{Y}} = 0$ and 1, and numerical results using $10^6$ mock data points for $r_{\tilde{X} \tilde{Y}} = 0.5$.}
    \label{fig:Dtdis}
\end{figure}

In the case of $\rho_p \gg \rho_e$, the individual timing residuals can be denoted as a product of two random variables, {\it i.e.},  
\begin{eqnarray}\label{eq:Dtpn1}
  \dt =c_1 X Y\,,
\end{eqnarray}
where $c_1 = \pi G\rho_p/(2m_a^3)$ is a uniform coefficient, and $\dt$, $X$ and $Y$ are shorthand notations for $\dt^a_{p,n}$ in Eq.~(\ref{eq:dti1}), $X_{p,n}^{(1)}$ in Eq.~(\ref{eq:Xpn}) and $Y_{p,n}^{(1)}$ in Eq.~(\ref{eq:Ypn}). Since $X$ and $Y$ are independent Gaussian variables, both having a zero mean and a unit variance, the composite variable $\dt$ is characterized by a probability distribution function (PDF) of variance gamma (VG) type:
\begin{equation}\label{eq:VGdis}
p(\dt|c_1)=\frac{1}{\pi \sigma }K_0\left( \frac{\left|\dt\right|}{\sigma} \right),\; \sigma^2=c_1\,,
\end{equation}
where $K_0(z)$ is the modified Bessel function of the second kind. We compare the VG distribution with a Gaussian distribution in the left panel of Fig.~\ref{fig:Dtdis}. The two distributions exhibit different behaviors. While being skewness-free, the VG distribution is characterized by a larger kurtosis than the Gaussian distribution. At small $z$, the VG distribution features a sharp peak, driven by the logarithmic divergence of the Bessel function near the origin, {\it i.e.},  
$p(\dt|c_1)\approx \frac{1}{\pi \sigma}\left[ -\ln (\dt/\sigma)+\ln 2-\gamma_{\rm E}\right]$.
Here $\gamma_{\rm E}\approx 0.5772$ is the Euler-Mascheroni constant. For large $z$, the VG distribution displays a longer tail, reflecting a slower decay of its PDF. It follows 
$p(\dt|c_1)\approx \frac{1}{\sqrt{2\pi \sigma}}|\dt|^{-\frac{1}{2}}\mathrm{e}^{-|\dt|/\sigma}$, as $\dt$ approaches infinity.
These characteristics highlight the non-Gaussian nature of the VG distribution.

For the case of $\rho_p\approx\rho_e\approx \rho_0$, the individual timing residuals can be expressed as a function of four random variables:  
\begin{eqnarray}\label{eq:dti3}
    \dt = c'_1 \sum_{\mathbf{v},\mathbf{v}'\in \Omega}\mC_{\mathbf{v}}\mC_{\mathbf{v}'}\sin(\theta_{\mathbf{v}}+\theta_{\mathbf{v}'})\cos (\bar\phi_{\mathbf{v}}+\bar\phi_{\mathbf{v}'}) 
    = 2 c'_1 \left(X'Y'-U'V' \right)\,,
\end{eqnarray}
where $c'_1\equiv \pi G\rho_0/(2m_a^3)$, $\theta_{\mathbf{v}}=\frac{1}{2}m_a(L_p+\mathbf{v}\cdot \mathbf{x}_p)$, $\bar\phi_{\mathbf{v}}=\frac{1}{2}m_a(2t_n-L_p-\mathbf{v}\cdot \mathbf{x}_p)+\phi_{\mathbf{v}}$ and the four statistical variables  
\begin{eqnarray}\label{eq:XYUV}
    &&X'\equiv \sum_{\mathbf{v}\in \Omega}{\mC_{\mathbf{v}}\cos\theta_{\mathbf{v}}\cos \bar\phi_{\mathbf{v}}},\quad
    Y'\equiv \sum_{\mathbf{v}\in \Omega}\mC_{\mathbf{v}}\sin\theta_{\mathbf{v}}\cos \bar\phi_{\mathbf{v}}\,,\nonumber\\
    &&U'\equiv \sum_{\mathbf{v}\in \Omega}{\mC_{\mathbf{v}}\cos\theta_{\mathbf{v}}\sin \bar\phi_{\mathbf{v}}},\quad
    V'\equiv \sum_{\mathbf{v}\in \Omega}\mC_{\mathbf{v}}\sin\theta_{\mathbf{v}}\sin \bar\phi_{\mathbf{v}}\,.
\end{eqnarray}
Following the same reasoning as in Eq.~(\ref{eq:Xpn}), one can find that these variables are Gaussian, each having a zero mean and variances given by 
\begin{eqnarray}\label{eq:covXYUV}
  &&\sigma_{X'}=\sigma_{U'}=\frac{1}{2}\left[ 1+\mathrm{e}^{-\frac{1}{4}y_{ep}^{2}}\cos \left( m_aL_p-m_a\mathbf{v}_{\odot}\cdot \mathbf{x}_p \right) \right]\nonumber\\
  &&\sigma_{Y'}=\sigma_{V'}=\frac{1}{2}\left[ 1-\mathrm{e}^{-\frac{1}{4}y_{ep}^{2}}\cos \left( m_aL_p-m_a\mathbf{v}_{\odot}\cdot \mathbf{x}_p \right) \right] \,.
\end{eqnarray}
Notably, $X'$, $Y'$ are independent of $U'$, $V'$ since they are constructed from two 
independent sets of random variables, {\it i.e.}, $\{\mathcal C_{\mathbf v}\cos\bar\phi_{\mathbf v}\}$ and $\{\mathcal C_{\mathbf v}\sin\bar\phi_{\mathbf v} \}$, respectively. However, $X'$ and $Y'$ are correlated, 
so as $U'$ and $V'$. Using the velocity distribution in Eq.~(\ref{eq: SHM}), we obtain their Pearson correlation coefficients
\begin{eqnarray}\label{eq:Pcorr}
r_{\tilde X \tilde Y}\equiv \frac{\langle \tilde X \tilde Y\rangle}{\sigma_{\tilde X}\sigma_{\tilde Y}} =\frac{\mathrm{e}^{-\frac{1}{4}y_{ep}^{2}}\sin \left( m_aL_p-m_a\mathbf{v}_{\odot}\cdot \mathbf{x}_p \right)}{\sqrt{1-\mathrm{e}^{-\frac{1}{2}y_{ep}^{2}}\cos \left( m_aL_p-m_a\mathbf{v}_{\odot}\cdot \mathbf{x}_p \right)}}\,,
\end{eqnarray}
where $\{\tilde X, \tilde Y\}$ denotes $\{X',Y'\}$  and $\{U',V'\}$. $r_{\tilde X\tilde Y}$ approaches a value of $\mathcal O(1)$ for $y_{ep} \ll 1$ but tends to be zero when $y_{ep} \gg 1$. This behavior reflects the dependence of the Pearson correlation on $y_{ep}$, one of the parameters characterizing spatial correlations of signals. 
The composite variable $x=2 c_1' \tilde X\tilde Y$ is skewed, and its PDF is given by~\cite{7579552}
\begin{equation}  \label{eq: 2XY PDF}
p(x|c'_1)=\frac{1}{2\pi c_1' \sigma _{\tilde X}\sigma _{\tilde Y}\sqrt{1-r_{\tilde X\tilde Y}^{2}}}\mathrm{e}^{\frac{r_{\tilde X\tilde Y}}{2 c_1' \sigma_{\tilde X}\sigma_{\tilde Y}\left(1-r_{\tilde X\tilde Y}^{2} \right)}x}K_0\left( \frac{\left| x \right|}{2 c_1' \sigma_{\tilde X}\sigma_{\tilde Y}\left( 1-r_{\tilde X\tilde Y}^{2} \right)} \right)\,.
\end{equation}

Determining the PDF of $\dt$ in Eq.~(\ref{eq:dti3}) remains a challenge, even with the known PDFs of both $2c_1'X' Y'$ and  $2c_1'U' V'$ in Eq.~(\ref{eq: 2XY PDF}) (see Appendix~\ref{app:1} for more discussions on the PDF in a general case). However, analytical expressions for the PDF can be derived in two $r_{\tilde X \tilde Y}$ limits:   
\begin{eqnarray} \label{eq:fDtlimit}
     && r_{\tilde X\tilde Y}=1:\; p(\dt|c'_1)=\frac{1}{\pi\sigma}K_0\left( \frac{\left| \dt \right|}{\sigma} \right) \, , \quad  \sigma=4 c_1'\sigma_{\tilde X}\sigma_{\tilde Y}   \, ; \\
     && r_{\tilde X\tilde Y}=0:\; p(\dt|c'_1)=\frac{1}{2b}\mathrm{e}^{-\frac{\left| \dt \right|}{b}} \, ,  \quad  \quad  \quad  \quad  b=2c_1'\sigma_{\tilde X}\sigma_{\tilde Y}\, .
\end{eqnarray}
Specifically, $\Delta t$ respects the VG distribution in the limit of $r_{\tilde X \tilde Y}=1$, as it occurs to the case of $\rho_p\gg \rho_e$, and the Laplace distribution in the limit of $r_{\tilde X \tilde Y}=0$. We demonstrate in the right panel of  Fig.~\ref{fig:Dtdis} the PDFs of $\Delta t$ for different $r_{\tilde X \tilde Y}$ values. As the $r_{\tilde X \tilde Y}$ decreases from one to zero, the peak becomes less sharp and the tail also becomes less long. Despite this feature, the Laplace distribution decays still more slowly in tail than the Gaussian distribution.  This comparison also reveals that the parameter $y_{eq}$ not only mediates spatial correlations of the ALDM timing signals shown in Eq.~(\ref{eq:Dtcorr}), but also affects statistical properties of these signals when both the ``pulsar'' and ``Earth'' terms are present.

The non-Gaussian statistics of the ALDM timing signals pose challenges for the construction of likelihood in Bayesian analysis. To estimate the applicability of Gaussian approximation, let us quantify the similarity between the aforementioned VG and Laplace distributions and a Gaussian distribution, using the method of series expansions. In cosmology and astronomy, the Gram-Charlier A series, Edgeworth series, and Gauss-Hermite polynomial series have been widely used for parametrizing signal or noise non-Gaussianity~\cite{Blinnikov:1997jq, Bartolo:2004if, Lentati:2014hja}. We take the Gauss-Hermite polynomial series $H_n(x)$ for demonstration, considering their good  convergence~\cite{vanderMarel:1993zz}. Then we expand the target PDF $p(x)$ 
in this orthonormal basis as 
\begin{equation}\label{eqn:wavefunctionExpansion}
   \sqrt{p(x)} 
   = \sum_{n=0}^\infty \alpha_n\exp\left(-\frac{x^2}{4\sigma^2}\right)C_nH_n\left(\frac{x}{\sqrt{2}\sigma}\right) \, ,
\end{equation}
where $\sigma$ is a variance parameter and  $C_n = (2^n n! \sqrt{2\pi}\sigma)^{-1/2}$ are normalization factors. The expansion coefficients  
\begin{equation}\label{eqn:expansionalpha} 
    \alpha_n\equiv C_n\int_{-\infty}^{+\infty} \sqrt{p(x)}\,
    H_n\left(\frac{x'}{\sqrt{2}\sigma}\right)\exp\left(-\frac{(x')^2}{4\sigma^2}\right) dx' 
\end{equation}
are real, and satisfy the normalization condition $\sum_{n=0}^\infty\alpha_n^2=1$. The zeroth-order term in Eq.~(\ref{eqn:wavefunctionExpansion}) corresponds to a Gaussian, while non-Gaussian corrections are provided by higher-order terms. To fit the target PDF, we use the $N$-th order truncation of the series, which is given by
\begin{equation}\label{eq:ANx}
    \mathrm{A}_N(x)\equiv\sum_{n=0}^N\frac{\alpha_n}{M}C_nH_n\left(\frac{x'}{\sqrt{2}\sigma}\right)\exp\left(-\frac{(x')^2}{4\sigma^2}\right)\,.
\end{equation}
Here, the coefficients $\alpha_n$ are replaced with $\alpha_n / M$,  where $M=\sqrt{\sum_{i=0}^N |\alpha_i|^2}$, to maintain the normalization condition and ensure its validity to describe a distribution.

\begin{figure}
    \centering
    \includegraphics[height=4.5cm]{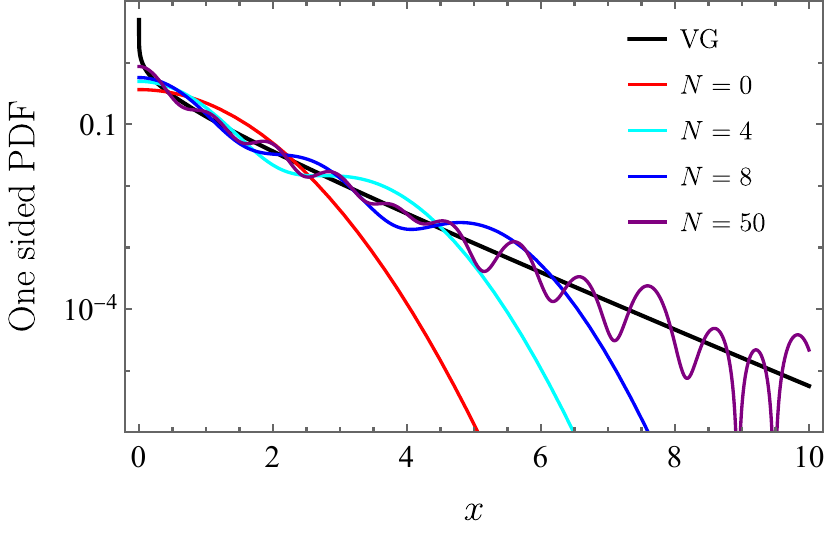}
    \includegraphics[height=4.5cm]{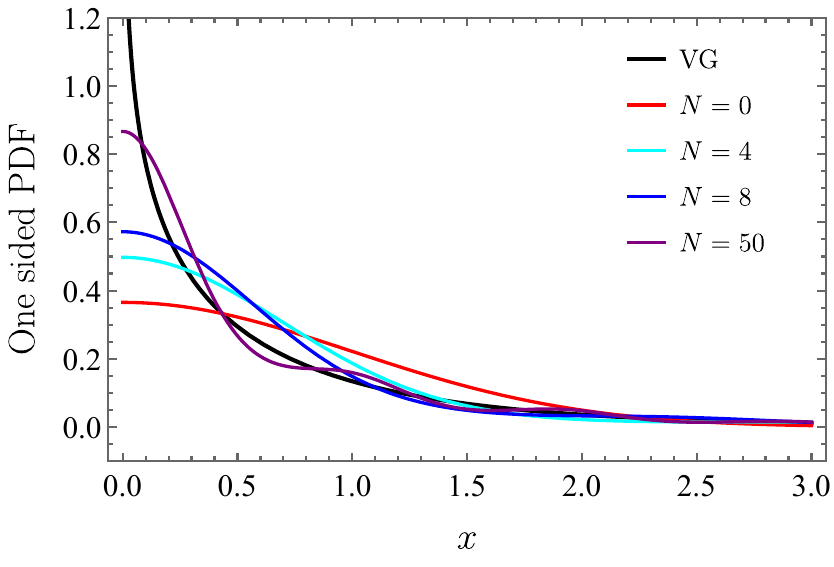}\\
    \includegraphics[height=4.5cm]{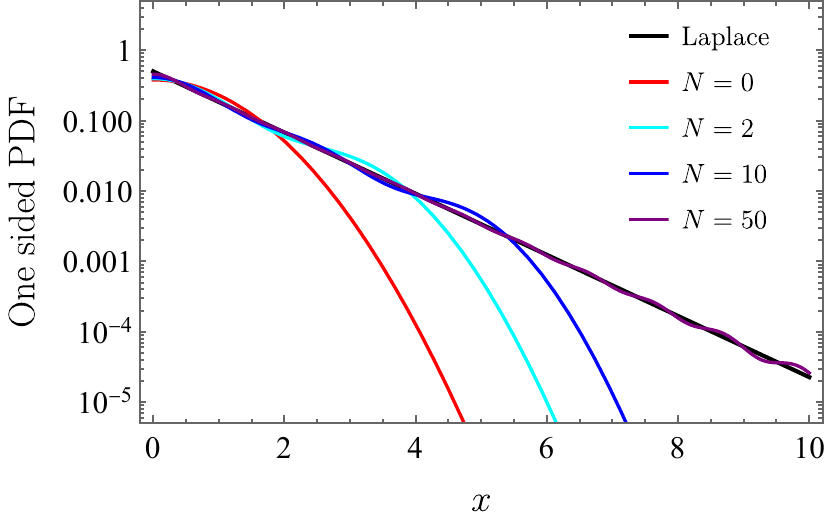}
    \includegraphics[height=4.5cm]{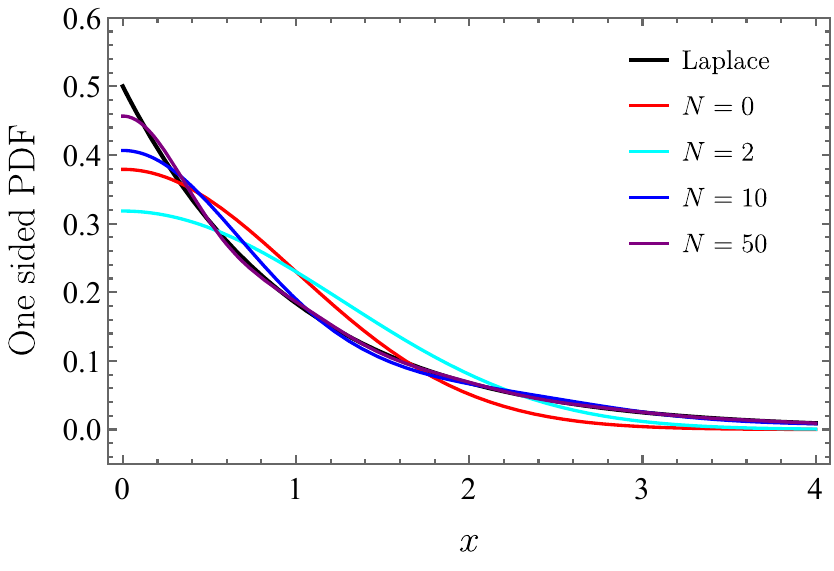}
    \captionsetup{justification=raggedright, singlelinecheck=false}
    \caption{Gauss-Hermite expansion of the VG and Laplace PDFs truncated at order $N$, shown with both logarithm (left) and linear (right) scales.}
    \label{fig:G-H expansion}
\end{figure}

\begin{table}[]
\begin{tabular}{c|l|lllllllll}
\hline
\multicolumn{1}{l|}{}    & $n, N$\quad     & 0    & 2      & 4     & 6      & 8     & 10     & $\cdots$ & 50      & 100    \\ \hline
\multirow{2}{*}{VG}      & $\alpha_n$ & 0.96\; & -0.059\; & 0.19\;  & -0.053\; & 0.098\; & -0.045\; & $\cdots$\; & -0.024\;  & 0.011\;  \\
                         & $H_N$      & 0.15 & 0.15   & 0.11  & 0.10   & 0.089 & 0.086  & $\cdots$ & 0.040   & 0.0051 \\ \hline
\multirow{2}{*}{Laplace} & $\alpha_n$ & 0.97 & 0.12   & 0.17  & 0.031  & 0.070 & 0.0098 & $\cdots$ & -0.0025 & 0.0013 \\
                         & $H_N$      & 0.11 & 0.097  & 0.047 & 0.044  & 0.027 & 0.027  & $\cdots$ & 0.0060  & 0.0036  \\ \hline
\end{tabular}
\captionsetup{justification=raggedright, singlelinecheck=false}
\caption{Gauss-Hermite expansion coefficients $\alpha_n$ in Eq.~(\ref{eqn:expansionalpha}) for the VG and Laplace distributions, and their Hellinger distances to the $N$-th order truncation $A_N$ in Eq.~(\ref{eq:ANx}). Since both distributions are symmetric, $\alpha_{2i+1}$ vanishes for $i\in \mathbb{N}$. }
\label{tab:alpha_n}
\end{table}

We demonstrate in Fig.~\ref{fig:G-H expansion} the Gauss-Hermite expansion of the VG and Laplace PDFs.
As shown in this figure, including higher-order corrections progressively reduces the difference from a Gaussian case on both peak and tail. We also calculate the expansion coefficients for the VG and Laplace distributions numerically and show them in Table \ref{tab:alpha_n}. Both cases demonstrate a good convergence as $n$ increases, with $\alpha_0$, namely the Gaussian component, yielding a contribution more than $90\%$ to $p(x)$.

To further quantify the convergence of this series expansion, we can leverage as a measure the Hellinger distance, defined
\begin{equation}
    H(P,Q)\equiv\left[\frac 12\int\left(\sqrt{p(x)}-\sqrt{q(x)}\right)^2\mathrm dx\right]^{1/2} 
\end{equation}
for distributions $P$ and $Q$. Here, $p(x)$ and $q(x)$ are the properly normalized PDFs of $P$ and $Q$. The Hellinger distance satisfies:
\begin{equation}
    0\leq H(P,Q)\leq 1,
\end{equation}
with $H(P,Q)=0$ only if $P$ and $Q$ are identical distributions. For our case, we take $\sqrt{p(x)}$ as for the target PDF, and $\sqrt{q(x)}=\mathrm{A}_{N}(x)$. Making use of their series expansions in Eqs.~(\ref{eqn:wavefunctionExpansion}) and (\ref{eq:ANx}), the Hellinger distance can be analytically derived as\footnote{To derive an analytical expression, we define $\sqrt{q(x)}=\mathrm{A}_{N}(x)$ instead of $\sqrt{q(x)}=|\mathrm{A}_{N}(x)|\geq0$, as required by the original definition. Since $|\sqrt{p}-\mathrm{A}_{N}(x)|\gtrsim  |\sqrt{p}-|\mathrm{A}_{N}(x)||$, the analytical result in Eq.~(\ref{eq:HN}) provides an upper bound on the Hellinger distance according to the original definition.}
\begin{equation}\label{eq:HN}
\begin{split}
    H_N={}&\frac{1}{\sqrt 2}\times\sqrt{M^2\left(1-\frac 1M\right)^2+1-M} \, .
\end{split}
\end{equation}
In Table~\ref{tab:alpha_n},  we calculate the Hellinger distance between the VG and Laplace distributions and their Gauss-Hermite truncations at order $N$.  In both cases, the Hellinger distance is small, with the expansion for the Laplace distribution converging more quickly than that of the VG distribution. This analysis offers a justification for using the Gaussian approximation for single ALDM-induced timing residual.

Extending the discussion from individual observations to multiple observations, represented by the signal vector
$\bDt^a$ in Eq.~(\ref{eq:Dtvec}), introduces additional complexities due to the challenge of obtaining the joint PDF $p(\bDt^a|\bDvth^a)$ for the timing signals, where $\bDvth^a$ denotes the ALDM parameters. The Gaussian approximation discussed above can greatly simplify this task, since a vector of Gaussian variables respects a multivariate Gaussian distribution. 
Yet, the method of series expansion cannot be straightforwardly applied without knowledge of $p(\bDt^a|\bDvth^a)$, and its applicability needs to be further examined. 

Alternatively, one can address this complexity by taking a  more generic treatment, leveraging the fact that the ALDM timing signals, while being non-Gaussian, arise from a construction of  Gaussian variables. Let us consider the case of $\rho_{p}\gg \rho_{e}$ as an example. In this case, the signal vector as a generalization of Eq.~(\ref{eq:Dtpn1}) is given by  
\begin{eqnarray}\label{eq:bDt1}
\bDt^a=c_1 \bD_X \bY\,,
\end{eqnarray}
where $\bX$ and $\bY$ are shorthands for $\bX^{(1)}$ and $\bY^{(1)}$, respectively, and $\bD_X$ is a matrix form of $\bX$, {\it i.e.},  $\bD_X={\rm diag}(\bX)$. 
The timing signal is a function of two Gaussian vectors, namely $\bX$ and $\bY$. To find the joint PDF of $\bDt^a$, one can take a Jacobian transformation from $\{\bX,\bY\}$ to $\{\bDt^a, \bY\}$, which yields
\begin{eqnarray}\label{eq:fbDta}
p(\bDt^a|\bDvth^a)=\int p\left(\{\bX(\bDt^a,\bY),\bY\}|\bDvth^a\right) |\det(\mathbf{J})| d\bY\, .
\end{eqnarray}
Here, $p(\{\bX,\bY\}|\bDvth^a)$ denotes the joint multivariate Gaussian PDF for $\bX$ and $\bY$, with its explicit form given in Eq.~(\ref{eq:fXY}) in Appendix~\ref{app:2}, and $\mathbf{J}$ is the Jacobian matrix. For individual observations, this calculation reproduces the VG distribution in Eq.~(\ref{eq:VGdis}). 
However, for multiple observations, the integration becomes very difficult due to high dimensionality of data. The situation could be even more involved in the case of $\rho_{p} \approx \rho_{e}$.  Therefore, rather than presenting a complete discussion, we will demonstrate in the next Subsection that this generic method can provide a consistency check in small-signal limit for the PTA likelihood calculated under the Gaussian approximation.

\subsection{PTA analysis scheme}
\label{sec:ptaanalysis}

The ALDM field $a(\mathbf{x},t)$ in our galaxy represents a specific realization of the nuisance parameters $\{\alpha_{\bf{v}}, \phi_{\bf{v}}\}$  in Eq.~(\ref{eq:axionf2}). For the ALDM mass range of interest, its signals however can only be probabilistically predicted, due to the statistical uncertainty of such a realization. To detect such signals in data, one method is to marginalize over the relevant random distributions of the signal vector~\cite{Centers:2019dyn}, as we have done for the PPA analysis of searching for the ultralight ALDM in~\cite{Liu:2021zlt,Xue:2024zjq}. Next, let us apply this method to the PTA analysis. 

Similar to the case of SGWB searches, we assume that the timing data contains the following contributions~\cite{NANOGrav:2023icp}
\begin{eqnarray}
\bDt=\bDt^{a}+\bDt_{\rm det}+\boldsymbol{F c}+\boldsymbol{n}\,,
\end{eqnarray}
where $\bDt^{a}$ is the ALDM-induced residuals (see Eq.~(\ref{eq:Dtvec})).  $\bDt_{\rm det}=\boldsymbol{M}\boldsymbol{\epsilon}$ represents the deterministic residuals from the timing model, where $\boldsymbol{M}$ is the design matrix, and $\boldsymbol{\epsilon}$ denotes the corresponding parameter offsets. $\boldsymbol{F c}$ denotes red noises with $\boldsymbol{F}$ the Fourier design matrix and  $\boldsymbol{c}$ the Fourier coefficients, and $\boldsymbol{n}$ is white noise.
Assuming the random noises to be Gaussian, we can model PTA data using the multivariate Gaussian likelihood~\cite{vanHaasteren:2008yh}: 
\begin{equation}\label{eq:Lm}
\begin{split}
    p(\bDt|\bDvth^n_t,\bDt^a)=&{}\frac 1{\sqrt{\det(2\pi\boldsymbol C^{n}_t)}} \exp\left[-\frac{1}{2}\big(\bDt-\bDt_{\rm det}-\bDt^a\big)^T(\bC^{n}_t)^{-1}\big(\bDt-\bDt_{\rm det}-\bDt^a\big)\right]\,,
\end{split}
\end{equation}
where $\bDvth^n_t$ and $\bC^{n}_t$ represent the parameters and covariance matrix for the random noises, respectively. The covariance matrix can be further expressed as $\bC^{n}_t=\boldsymbol{N}+\boldsymbol{F^T \phi F}$, where $\boldsymbol{N}$ accounts for the white noise contribution and $\bphi$ denotes the frequency domain covariance for the red noise. 
As discussed previously, ALDM-induced residuals $\bDt^a$ can be treated as a stochastic non-Gaussian signal, characterized by the joint PDF $p(\bDt^a|\bDvth^a)$. Marginalizing the likelihood function over this PDF leads to
\begin{eqnarray}\label{eq:LmDt}
    p(\bDt|\bDvth_t)= \frac 1{\sqrt{\det(2\pi\boldsymbol C^{n}_t)}} \int \exp\left[-\frac{1}{2}\big(\bDt_o-\bDt^a\big)^T(\bC^{n}_t)^{-1}\big(\bDt_o-\bDt^a\big)\right] p(\bDt^a|\bDvth^a) d\bDt^a\,,
\end{eqnarray}
where $\bDvth_t=\{\bDvth_t^n,\bDvth^a\}$ and $\bDt_o\equiv \bDt- \bDt_{\rm det}$.

The exact calculation of $p(\bDt|\bDvth_t)$ is  challenging due to the difficulty in obtaining the joint PDF $p(\bDt^a|\bDvth^a)$ for multiple observations, such as in Eq.~(\ref{eq:fbDta}), as well as in performing the integration in Eq.~(\ref{eq:LmDt}). Building on the justification in Sec.~\ref{sec:Dtstat} for using a Gaussian approximation to model the PDF of individual ALDM-induced timing residuals, we first assume, as a proof of concept, that this Gaussian approximation can be generalized to the multi-dimensional case $\bDt^a$. This allows us to define a joint PDF of multivariate Gaussian 
\begin{eqnarray}
    p(\bDt^a|\bDvth^a)
    \approx \frac 1{\sqrt{\det(2\pi\bC^{a}_t)}}\exp\left[-\frac{1}{2}(\bDt^a)^T (\bC^{a}_t)^{-1}\bDt^a\right]\,,
\end{eqnarray}
with the signal covariance matrix $\bC^{a}_t=\langle \bDt^a (\bDt^a)^T\rangle$ (see  Eq.~(\ref{eq:Ctafull})). Under this approximation, the ALDM timing signal becomes a stochastic Gaussian process, allowing us to apply the same marginalization procedure as for other stochastic Gaussian signals in Eq.~(\ref{eq:Lm}) to Eq.~(\ref{eq:LmDt}), which yields
\begin{eqnarray}\label{eq:LmGaussian}
p^{(g)}(\bDt|\bDvth_t)
= \frac 1{\sqrt{\det(2\pi \bC_t)}} \exp\left[-\frac{1}{2}\bDt_o^T \bC^{-1}_t\bDt_o\right]\,,
\end{eqnarray}
where $\bC_t=\bC^{n}_t+\bC^{a}_t$ is the full covariance matrix. 

To further justify this approximation, we alternatively compute the marginalized likelihood $p(\bDt|\bDvth_t)  $ by directly marginalizing over the Gaussian variables (i.e. $X_{p,n}^{(i)}$ and $Y_{p,n}^{(i)}$ in Eqs.~(\ref{eq:Xpn}) and (\ref{eq:Ypn})) that make up $\bDt^a$. As a demonstration, let us consider the scenario where $\rho_p\gg\rho_e$. Using the signal defined with Eq.~(\ref{eq:bDt1}), the marginalized likelihood in Eq.~(\ref{eq:LmDt}) can be expressed as an integral over  $\bX$ and $\bY$ with their joint PDF $p(\{\bX,\bY\}|\bDvth^a)$ in Eq.~(\ref{eq:fbDta}), i.e. 
\begin{eqnarray}\label{eq:eq:LmDtXY}
p(\bDt|\bDvth_t) = \frac 1{\sqrt{\det(2\pi\boldsymbol C^{n}_t)}} \int\exp\left[-\frac{1}{2}\big(\bDt_o-c_1 \bD_X \bY\big)^T(\bC^{n}_t)^{-1}\big(\bDt_o-c_1 \bD_X \bY\big)\right]
p(\{\bX,\bY\}|\bDvth^a) d\bX d\bY .
\nonumber\\
\end{eqnarray}
The standard Gaussian integral over $\bY$ can be performed analytically. However, the variable $\bX$ cannot be marginalized in the same manner. Instead, we examine the small-signal limit by expanding with respect to the signal strength coefficient $c_1$. For comparison with the Gaussian approximation $p^{(g)}$ in Eq.~(\ref{eq:LmGaussian}), we focus on the exponential term that is quadratic in $\bDt_o$ for demonstration purposes. As detailed in Appendix~\ref{app:2}, the marginalized likelihood is then approximately given by
\begin{eqnarray}\label{eq:LmDtvec2}
    p(\bDt|\bDvth_t)
    &\propto& \exp\left[-\frac{1}{2}\bDt_o^T(\bC^{n}_t)^{-1}\bDt_o \right]\bigg[1+\frac{1}{2}c_1^2\tr(\bC_X \bD_t\bS \bD_t) + \frac{1}{2}c_1^2\Big(\tr(\bC_{XY}\bD_t\bC_{XY}\bD_t)\nonumber\\
    &&-\tr(\bC_X\bD_t\bC_{XY}\bC_X^{-1}\bC_{XY}\bD_t)\Big) +\mathcal{O}(c_1^3)\bigg]\nonumber\\
    &\propto& \exp\left[-\frac{1}{2}\bDt_o^T(\bC^{n}_t)^{-1}\bDt_o \right] \bigg[1+\frac{1}{2}\bDt_o^T(\bC^{n}_t)^{-1}\bC_t^{a}(\bC^{n}_t)^{-1}\bDt_o +\mathcal{O}(c_1^3)\bigg]\, ,
\end{eqnarray}
where $\bD_t={\rm diag}((\bC_t^{n})^{-1}\bDt_o)$.
In the last line we have used the signal covariance matrix $\bC^{a}_t$ in Eq.~(\ref{eq:Ctafull}) for the $\rho_p\gg\rho_e$ case. As expected, the leading correction arises at the order of $\mathcal O(c_1^2)$. 

This result for $p(\bDt|\bDvth_t)$ aligns perfectly with Eq.~(\ref{eq:LmGaussian}) in the small signal limit, {\it i.e.},  
\begin{eqnarray}\label{eq:LmGaussianexp}
 p^{(g)}(\bDt|\bDvth_t)
\propto \exp\left[-\frac{1}{2}\bDt_o^T (\bC^{n}_t)^{-1}\bDt_o\right]\left[1+\frac{1}{2}\bDt_o^T(\bC^{n}_t)^{-1}\bC^{a}_t(\bC^{n}_t)^{-1}\bDt_o+\mathcal{O}(c_1^3)\right]\,,
\end{eqnarray}
validating the Gaussian construction 
in Eq.~(\ref{eq:LmGaussian}) from a different perspective. 
Such consistency is expected for the case of random Gaussian noise. Given that the likelihood in Eq.~(\ref{eq:Lm}) describes the probability of observing the data given the model, the statistical properties of the data are primarily driven by noise in the small signal limit. Consequently, two-point correlation functions dominate over higher-point ones, and the non-Gaussianity of the signal is relevant only at subleading orders in the Bayesian analysis. Therefore, a Gaussian construction of the likelihood function in the small-signal limit is permissible, even if the signal is not Gaussian-like. For the case of $\rho_e \approx \rho_p$, which applies to existing PTAs, a similar form for $p(\bDt|\bDvth_t)$ is anticipated in the small-signal limit, though directly deriving it is more challenging. This reinforces the Gaussian approximation used in \cite{Luu:2023rgg} to search for the correlations of the ALDM timing signal. 


One technical challenge of using the Gaussian likelihood $p^{(g)}(\bDt|\bDvth_t)$ in Eq.~(\ref{eq:LmGaussian}) to perform  Bayesian analysis is accurately calculating the inverse of the covariance matrix. Given their high dimensionality and intricate structure, matrix decomposition techniques are often employed in detecting the nano-Hz SGWB, to streamline this process and enhance the stability of the results~\cite{NANOGrav:2023icp}. Next, let us consider their application for calculating $\bC_t^{-1}=(\boldsymbol{N}+\bF^T \bphi \bF+ \bC^a_t)^{-1}$ in  Eq.~(\ref{eq:LmGaussian}) (regarding their application in the PPA analysis, see~\cite{Xue:2024zjq}).
The white noise is uncorrelated across the epochs of all pulsars, making it straightforward to invert $\boldsymbol{N}$. In contrast, the red noise displays specific temporal trends and can be correlated among pulsars. Therefore, its covariance matrix is typically parametrized as $\bF^T \bphi \bF$, where the Fourier design matrix and the frequency domain covariance matrix are expressed as 
\begin{eqnarray}\label{eq:FPhir}
   \bF= \begin{pmatrix}
        \cos(2\pi f_1 t)\\
        \sin(2\pi f_1 t)\\
        ...\\
        \cos(2\pi f_{k_{\rm max}} t)\\
        \sin(2\pi f_{k_{\rm max}} t)
    \end{pmatrix},\quad
    \bphi= 
    \begin{pmatrix}
        \bphi_1\boldsymbol{I}_{2\times 2} &  & & \\ &\bphi_2\boldsymbol{I}_{2\times 2}  & & \\
        & & \ddots & \\ 
        & & & \bphi_{k_{\rm max}}\boldsymbol{I}_{2\times 2}\\
    \end{pmatrix}\,.
\end{eqnarray}
Here, $f_k=k/T_p$ represents the $k$-th frequency bin, $T_p$ is the observation time span, and $k_{\rm max}$ denotes the total number of frequency bins. The covariance matrix $\bphi$ is diagonal in frequency space, and each frequency bin component, $\bphi_{k}$, can be decomposed into two parts: $\bphi_{k,pq}=\delta_{pq}\varphi_{pk}+\Gamma_{pq}\Phi_k$. The first part represents intrinsic red noise that is uncorrelated among pulsars, with $\varphi_{pk}$ encoding spectral information. The second term denotes correlated red noise,  where $\Gamma_{pq}$ represents the characteristic correlation pattern  among pulsars and $\Phi_k$ provides the spectral information. 

For the ALDM timing signals, the covariance matrix in Eq.~\eqref{eq:Ctafull} can be similarly decomposed as $\bC^a_t = (\bF^a)^T \bphi^a \bF^a$, with
\begin{equation}
    \boldsymbol{F}^a=
    \begin{pmatrix}
        \cos(2m_a t)\\
        \sin(2m_a t)
    \end{pmatrix},\quad
    \bphi^a=
    \begin{pmatrix}
        \bphi^a_{cc}&\bphi^a_{cs}\\
        \bphi^a_{sc}&\bphi^a_{ss}\\
    \end{pmatrix}\,.
\end{equation}
The matrix $\bphi^a$ is no longer diagonal, with a nonzero off-diagonal term for the two components of $\bF^a$. Its components, which encode essential information about temporal and spatial correlations of the ALDM signal, can be expressed as
\begin{eqnarray}\label{eq:Phipqcomp} \bphi^a_{cc,pq}
&=&\frac{\pi^2G^2}{4m_a^6}\bigg\{\rho_e^2+\rho_p\rho_q\cos\left[2m_a( L_{pq}-\mathbf{v}_{\odot}\cdot  \mathbf{x}_{pq})\right]e^{-\frac{1}{2}y_{pq}^2}-\rho_e\rho_p\cos\left[2m_a(L_p-\mathbf{v}_{\odot}\cdot  \mathbf{x}_{pe})\right]e^{-\frac{1}{2}y_{pe}^2}\nonumber\\
&&-\rho_e\rho_q\cos\left[2m_a(L_q-\mathbf{v}_{\odot}\cdot  \mathbf{x}_{qe})\right]e^{-\frac{1}{2}y_{qe}^2}\bigg\}   \, , \nonumber\\
\bphi^a_{sc,pq}
&=&\frac{\pi^2G^2}{4m_a^6}\bigg\{ \rho_p\rho_q\sin\left[2m_a(L_{pq}-\mathbf{v}_{\odot}\cdot  \mathbf{x}_{pq})\right]e^{-\frac{1}{2}y_{pq}^2}-\rho_e\rho_p\sin\left[2m_a(L_p-\mathbf{v}_{\odot}\cdot  \mathbf{x}_{pe})\right]e^{-\frac{1}{2}y_{pe}^2}\nonumber\\
&&+\rho_e\rho_q\sin\left[2m_a(L_q-\mathbf{v}_{\odot}\cdot  \mathbf{x}_{qe})\right]e^{-\frac{1}{2}y_{qe}^2}
\bigg\}\, ,
\end{eqnarray}
and $\bphi^a_{ss}=\bphi^a_{cc}$ and $\bphi^a_{cs}=-\bphi^a_{sc}$.

Following the method used for the PPA analysis in~\cite{Xue:2024zjq}, we then define   
\begin{equation}
 \tilde{\bphi}=\mathrm{diag}\{\bphi,\bphi^a\},\quad
    \tilde\bF=
    \begin{pmatrix}
        \bF\\
        \bF^a
    \end{pmatrix}\,,
\end{equation}
such that the red noise and the ALDM signal can be combined into a single term:
\begin{equation}
    \boldsymbol{F^T \phi F}+\boldsymbol{C}^a_t=\tilde\bF^T\tilde{\boldsymbol\phi}\tilde\bF\, .
\end{equation}
Applying the Woodbury matrix identity, the inverse of the total covariance matrix reads 
\begin{equation}
\boldsymbol C_t^{-1} = 
\left(\boldsymbol{N} + \tilde\bF^T \tilde{\bphi} \tilde\bF \right)^{-1}
= \boldsymbol{N}^{-1} - \boldsymbol{N}^{-1} \tilde\bF^T 
\left( \tilde{\boldsymbol\phi}^{-1} + \tilde\bF \boldsymbol{N}^{-1} \tilde\bF^T \right)^{-1} 
\tilde\bF \boldsymbol{N}^{-1}.
\end{equation}

We can now make a comparison between the SGWB and ALDM PTA signals. First, the ALDM-induced timing residuals  oscillate approximately with a single frequency which is determined by the ALDM mass, {\it i.e.},  $2\pi f\approx 2m_a$. In contrast, the SGWB signals are often modeled as a power-law spectrum, varying over a certain frequency range. Second, for $m_a \sim 10^{-22}\,$eV or equivalently $f\sim 10^{-9}\,$Hz, the de Broglie wavelength is $\sim \mathcal O(1000)\,$pc for the ALDM but only $\sim \mathcal O(1)\,$pc for the SGWB. This difference arises from  that the SGWB is relativistic while the ALDM is nonrelativistic. Since the distance of the arrayed pulsars to the Earth and their mutual distance usually vary from hundreds of to thousands of parsecs, the spatial correlations encoded in the exponential factors in Eq.~(\ref{eq:Phipqcomp}) can be significant and even further enhanced if the ALDM halo surrounding the pulsars is dense, but their counterparts for the SGWB signals get exponentially suppressed for the currently constructed PTAs (see discussions in Sec.~\ref{subsec:PTAcf} also). With these subleading contributions neglected, the SGWB covariance matrix in frequency domain  shares a structure of $\Gamma_{pq}\Phi_{k}$ in Eq.~(\ref{eq:FPhir}), with the corresponding $\Gamma_{pq}$ leading to the well-known Hellings–Downs curve. This explains why the Hellings–Downs correlation does not depend on the distance parameters of pulsars, but on their angular separations only. Finally, the covariance matrix for SGWBs is diagonal in the frequency domain, while $\bphi^a$ for ALDM shows a distinct structure with a similar Fourier design matrix. These differences highlight the unique properties of the ALDM PTA signals.

The inclusion of pulsar terms in the covariance matrix necessitates precise knowledge of pulsar location.  Yet, measuring pulsar distance is challenging, with relative uncertainties typically $\sim \mathcal O(10)\%$ or smaller.   The uncertainties of pulsar distance affect the covariance matrix through trigonometric and exponential functions, which are characterized by the ALDM Compton wavelength $\sim 1/m_a$ (where the solar motion phase term is negligible) and de Broglie wavelength $l_c\sim 1/m_a v_0$, respectively. In the analysis, they can be marginalized using the priors  determined by the measurement methods of pulsar distance, as done in~\cite{Luu:2023rgg,Xue:2024zjq}.
Although these uncertainties, generally much larger than $1/m_a$, would average out the trigonometric functions to zero in marginalization, the exponential dependency of the marginalized likelihood $p^{(g)}(\bDt|\bDvth_t)$ in Eq.~(\ref{eq:LmGaussian}) on $\bC^a_t$ tends to soften this effect. As detailed in Ref. \cite{Luu:2023rgg}, the information of spatial correlations encoded in sinc functions (or equivalently the exponential factors in Eq.~(\ref{eq:Phipqcomp})) remains largely intact unless the distance uncertainties exceed $l_c$. Therefore, including  pulsar terms and considering the influence of distance uncertainties are essential for the PTA analysis.

\subsection{PTA-PPA analysis scheme}
\label{sec:ptappaanalysis}

In contrast to the PTA case, where the signal two-point correlation function plays a crucial role, constructing the likelihood function for the PTA-PPA analysis is more challenging. This is because the two-point correlation function between the ALDM timing and polarization signals vanishes, and the leading-order statistical effect arises from their three-point correlation function (see Eq.~(\ref{eq:DPADPADtcorr})). Below, we will conduct an exploratory study to tackle this task, utilizing an approximate Gaussian construction along with a more fundamental treatment of the signal based on its Gaussian components.

For the first case, let us define the data vector as   
\begin{equation}\label{eq:Vo1}
 \bV_o = \left (\bDt_o ,  \textrm{vec}\{  \bDPA_o\bDPA_o^T -\langle\bDPA_o\bDPA_o^T \rangle   \}  \right )^T   \, ,
\end{equation}
where $\bDt_o$ is defined as in Eq.~(\ref{eq:LmDt}) and $\bDPA_o\equiv \bDPA- \bDPA_{\rm det}$ represents PA residuals with the deterministic noise subtracted. Using the model of PA residuals in~\cite{Xue:2024zjq}, we have $\bDPA_o=\bDPA^{n}+\bDPA^{a}$. 
In the quadratic $\bDPA_o\bDPA_o^T$, the ensemble mean has been subtracted to ensure that $\bV_o$ has a zero mean. vec$\{ \cdot \}$ denotes an operation of matrix vectorization. The diagonal elements involve squares and might be highly non-Gaussian, so they have been excluded. Assuming no correlation between $\bDPA^{n}$ and $\bDPA^a$, {\it i.e.}, $\langle\bDPA^{n}\bDPA^a\rangle=\langle\bDPA^{n}\rangle\langle\bDPA^a\rangle=0$, we have $\langle\bDPA_o\bDPA_o\rangle=\langle\bDPA^{n}\bDPA^{n}\rangle+\langle\bDPA^a\bDPA^a\rangle$.

Yet, at higher order there could exist contributions from a crossing between $\bDPA^{n}$ and $\bDPA^a$ in the polarization data constructed in Eq.~(\ref{eq:Vo1}).  So, instead of constructing the marginalized likelihood with a signal prior, as in Eq.~(\ref{eq:LmDt}),  we assume that the data vector $\bV_o$ follows a multivariate Gaussian distribution, which could be driven by large random Gaussian noise, and build an approximate Gaussian likelihood analogous to that for $\bDt_o$ (see Eq.~(\ref{eq:LmGaussian})). This leads to 
\begin{equation} \label{eq:tp1}
p^{(g)}(\bV|\bDvth)
=\frac{1}{{\sqrt{\det(2\pi \boldsymbol{K})}}}\exp\left[-\frac{1}{2}\bV_o^T \boldsymbol{K}^{-1}\bV_o\right]\,.
\end{equation}
Here, $\bDvth$ denotes all relevant parameters for both the ALDM signal and the noise. $\boldsymbol K$ is the full covariance matrix and can be written in block partitions:
\begin{equation}
    \boldsymbol K
    =
    \begin{pmatrix}
        \boldsymbol C_t
        &\left(\boldsymbol C_{\Pi t}\right)^T
        \\
        \boldsymbol C_{\Pi t}
        &\boldsymbol C_\Pi
    \end{pmatrix}  \, .
\end{equation}
Here, $\bC_t$ is the covariance matrix for $\bDt_o$ as in  Eq.~(\ref{eq:LmGaussian}), $\bC_\Pi$ corresponds to the four-point correlation functions for $\bDPA_o$, and the off-diagonal block $\bC_{\Pi t}$ encodes the correlations between $\bDt_o$ and $\bDPA_o$.  
Specifically, the $\boldsymbol C_{\Pi t}$ entries are given by
\begin{equation}  \label{eq:Cpit}
\begin{split}
    \left(\boldsymbol C_{\Pi t}\right)_{pn,qm;rl}
    = {}&\langle \dPA_{p,n}^{n} \dPA_{q,m}^{n} \dt_{r,l}^{n}\rangle + \langle \dPA_{p,n}^a \dPA_{q,m}^a \dt_{r,l}^a\rangle + ... ... \, .
\end{split}
\end{equation}
The second term is exactly the three-point correlation function between the ALDM-induced timing and polarization residuals given in 
Eq.~(\ref{eq:DPADPADtcorr}). The first term reflects the potential correlation of random noise in the timing and PA residuals.

Next, let us consider the alternative treatment, by taking Gaussianity of the ALDM random field as an input. We define the data vector as $\bV=(\bDt,\bDPA)^T$ and 
\begin{equation}\label{eq:Vo2}
 \boldsymbol{V}_o = \left(\bDt_o ,   \bDPA_o  \right)^T   \, ,
\end{equation}
using the same $\bDt_o$ and $\bDPA_o$ as in Eq.~(\ref{eq:Vo1}). For simplicity, we first assume the random noises for timing and polarization data to be independent. Consequently, the likelihood for the PTA-PPA analysis can be written as 
\begin{eqnarray}\label{eq:LDtDPA}
p(\bV|\bDvth^n,\bV^a)&=&p(\bDt|\bDvth^n_t,\bDt^a)p(\bDPA|\bDvth^n_{\PA},\bDPA^a)\nonumber\\
   &\propto& \exp\left[-\frac{1}{2}\big(\bDt_o-\bDt^a\big)^T(\bC^{n}_t)^{-1}\big(\bDt_o-\bDt^a\big)\right]\nonumber\\
   &\times& \exp\left[-\frac{1}{2}\big(\bDPA_o-\bDPA^a\big)^T(\bC^{n}_{\PA})^{-1}\big(\bDPA_o-\bDPA^a\big)\right]\,,
\end{eqnarray}
where $\bDvth^n=\{\bDvth^n_t,\bDvth^n_{\PA}\}$ and   $\bV^a= (\bDt^a,\,\bDPA^a)^T$. $\bDvth^n_{\PA}$ and $\bC_{\rm PA}^{n}$ represent the parameters and covariance matrix for random noise of PA residuals, as modeled in~\cite{Xue:2024zjq}. The next step is to marginalize over the correlated timing signal $\bDt^a$ and polarization signal $\bDPA^a$,  using their joint PDF.

As for the case of timing signal, deriving the exact form of the marginalized likelihood for the combined analysis from Eq.~(\ref{eq:LDtDPA}) is challenging due to the unknown  joint PDF of $\bDt^a$. Instead, we focus on deriving the marginalized likelihood by integrating over the Gaussian variables that constitute the timing and polarization signals. For demonstration purposes, we again consider the scenario where $\rho_p \gg \rho_e$. In this limit, the ALDM signal vectors can be expressed as
\begin{eqnarray}
\bDt^a=c_1 \bD_X \bY, \quad 
\bDPA^a=c_2 \bX\,,
\end{eqnarray}
where $\bX$ and $\bY$ are Gaussian variables defined in Eq.~(\ref{eq:bDt1}), with  $c_2=g_{a\gamma\gamma}\sqrt{\rho_p}/m_a$.
Similarly, the marginalized likelihood for the combined analysis can be derived by marginalizing over $\bV^a$ in Eq.~(\ref{eq:LDtDPA}) with the joint PDF $p(\{\bX,\bY\}|\bDvth^a)$
from Eq.~(\ref{eq:fbDta}),  
\begin{eqnarray}
p(\bV|\bDvth)
&=& \frac {1}{\sqrt{\det(2\pi \bC^{n}_t)\det(2\pi \bC^{n}_{\PA})}} \int\exp\left[-\frac{1}{2}\big(\bDt_o-c_1 \bD_X \bY\big)^T(\bC^{n}_t)^{-1}\big(\bDt_o-c_1 \bD_X \bY\big)\right]\nonumber\\
&&\times \exp\left[-\frac{1}{2}\big(\bDPA_o-c_2 \bX\big)^T(\bC^{n}_{\PA})^{-1}\big(\bDPA_o-c_2 \bX\big)\right]p(\{\bX,\bY\}|\bDvth^a) d\bX d\bY \,,
\end{eqnarray}
where $\bDvth=\{\bDvth^n,\bDvth^a\}$.
By performing the integration over these variables in the small-signal limit, we find that the marginalized likelihood is approximately given by (see Appendix~\ref{app:2} for details):
\begin{eqnarray} \label{eq:TPL}
p(\bV|\bDvth)&\propto&
\exp\left[-\frac{1}{2}\left(\bDt_o^T(\bC_{t}^{n})^{-1}\bDt_o+\bDPA_o^T(\bC^{n}_{\PA})^{-1}\bDPA_o\right) \right]\nonumber\\
    && \times \exp\bigg[\frac{1}{2}\bDPA_o^T(\bC^{n}_{\PA})^{-1}\bC^a_\PA(\bC^{n}_{\PA})^{-1}\bDPA_o\bigg]\bigg[1+\frac{1}{2}\bDt_o^T(\bC^{n}_t)^{-1}\bC_t^{a}(\bC^{n}_t)^{-1}\bDt_o\bigg]\nonumber\\
    && \times \exp\bigg[\frac{1}{2}c_1c_2^2\bDPA_o^T(\bC^{n}_{\PA})^{-1}(\bC_X\bD_t\bC_{XY}-\bC_{XY}\bD_t\bC_X)(\bC^{n}_{\PA})^{-1}\bDPA_o\bigg] \, .
\end{eqnarray}
Here, we have used the explicit form of signal covariance matrices $\bC_\PA^{a}$ and $\bC_t^{a}$ in Eqs.~(\ref{eq:CPAa}) and ~(\ref{eq:Ctafull}) for the $\rho_p\gg\rho_e$ case. 
The second line then encodes the two-point correlations of $\bDPA^a$, and the correlations of $\bDt^a$ in the small-signal limit, as provided in Eq.~(\ref{eq:LmDtvec2}). The third line can be rearranged as
\begin{eqnarray}\label{eq:LmDPADtcorre}
    &&\bDPA_o^T(\bC^{n}_{\PA})^{-1}(\bC_X\bD_t\bC_{XY}-\bC_{XY}\bD_t\bC_X)(\bC^{n}_{\PA})^{-1}\bDPA_o\nonumber\\
    &=&\bV_{\rm PA,n}\bV_{\rm PA,m}\bV_{\rm t,l}\left[(\bC_X)_{nl}(\bC_{XY})_{ml}+(\bC_X)_{ml}(\bC_{XY})_{nl}\right]\,,
\end{eqnarray}
where $\bV_{\rm PA}=(\bC_{\rm PA}^{n})^{-1}\bDPA_o$ and $\bV_{t}=(\bC_{\rm PA}^{n})^{-1}\bDt_o$. The three-point function in Eq.~(\ref{eq:DPADPADtcorr}) is thus revealed, since
\begin{eqnarray}
    \langle  \bDPA^a_{n}\bDPA^a_{m}\bDt^a_{l}\rangle=c_1 c_2^2 \langle \bX_n\bX_m \bX_l \bY_l\rangle
    =c_1 c_2^2 \left[(\bC_X)_{nl}(\bC_{XY})_{ml}+(\bC_X)_{ml}(\bC_{XY})_{nl}\right]\,.
\end{eqnarray}
This consistency between the leading-order correction to the marginalized likelihood and the leading-order correlation function is again attributed to the dominance of Gaussian noise in the small signal limit, similar to the timing signal case. 

This derivation can be extended to include potential correlations between random timing and polarization noises. For this case, it is reasonable to assume that the random noise vector $\bV^n=\bV_o-\bV^a$ is described by a multivariate Gaussian distribution with zero mean and relevant correlations. Eq.~(\ref{eq:LDtDPA}) can be then rewritten as
\begin{eqnarray}
 p(\bV|\bDvth^n,\bV^a)
\propto \exp\left[-\frac{1}{2}\big(\bV_o-\bV^a\big)^T(\bC^{n}_V)^{-1}\big(\bV_o-\bV^a\big)\right]\,,
\end{eqnarray}
where $\bC^{n}_V$ is the full covariance matrix for the noise vector $\bV^n$. The possible correlation between timing and polarization noises is encoded in its off-diagonal blocks, namely $\langle \bDt^{n} (\bDPA^{n})^T \rangle$, leading to additional contributions to the marginalized likelihood $p(\bV|\bDvth)$ in Eq.~(\ref{eq:TPL}). Note that this calculation is not expected to reproduce the noise correlation counterpart in Eq.~(\ref{eq:tp1}), as different statistical properties have been assumed for the noise. Finally, it is important to  extend this exploration to the  more realistic case where $\rho_e \approx \rho_p$, which however is quite involved since the marginalization of the ALDM signals requires an integration over four random Gaussian variables instead. We reserve these studies for future research.

While a complete likelihood for full Bayesian inference is not yet available, qualitative impacts of the cross‑correlation term on the inference of ALDM parameters can be seen. As Eq.~(\ref{eq:TPL}) shows, the combined likelihood contains three contributions: a PPA‑only term, a PTA‑only term, and a PTA‑PPA cross‑correlation term. In the absence of the PTA‑PPA term, the polarization signal and timing signal amplitudes (i.e. $c_2$ and $c_1$) can be constrained independently, yielding separate allowed bands in the plane of ALDM energy density and Chern‑Simons coupling. Including the cross‑correlation term, however, couples the two amplitudes, thereby turning the separate bands into a closed contour and constraining the ALDM parameters more tightly.

\section{Summary}
\label{sec:summary}

The ultralight ALDM, as one of the most representative DM candidates, predicts a strong wave nature on astronomical scales, which may leave distinct patterns in pulsar timing and polarization data and can be efficiently detected using PTA and PPA. Interestingly, the signal timing and polarization residuals arise from the ALDM gravitational perturbations to galactic metric and its nongravitational CB effects for photons. These two methods thus can be further combined to synergistically enhance the pulsar array’s capability to identify the signals. In this paper, we systematically explore characteristic correlation patterns of the ALDM  polarization and timing signals, investigate their statistical properties, and explore the construction of relevant likelihood functions in Bayesian analysis framework. 

In Sec.~\ref{sec:correlation}, we first revisit the previously derived  ALDM two-point correlation functions for PTA and PPA, and then extend the analysis to include correlations between its timing and polarization signals. The ALDM halo as a superposition of numerous particle plane waves exhibits a stochastic nature and can be effectively described as a random Gaussian field $a(\mathbf{x},t)$. The ALDM-induced PA residual shows a linear dependence on $a(\mathbf{x},t)$, making itself a random Gaussian variable. Consequently, the signal vector for the PPA follows a multivariate Gaussian distribution, with its statistical information completely encoded in its two-point correlation function. In contrast, the ALDM-induced timing residual depends on $a(\mathbf{x},t)$ quadratically, rendering itself non-Gaussian. Its statistical properties at leading order are  encoded in its two-point correlation function, which is essentially a four-point correlation function of $a(\mathbf{x},t)$.  Thus, the ALDM timing signal demonstrates distinct correlation patterns from those of its polarization signal. 
In relation to this non-Gaussianity also, the cross-correlation between the ALDM timing and polarization signals at leading order arises from their three-point correlation function, where two PA residuals and one timing residual interplay. The two-point cross-correlation vanishes since it is essentially a three-point correlation function of $a(\mathbf{x},t)$.  The spatial correlations of the ALDM signals across pulsars are characterized by an exponential factor $\sim e^{-\frac{1}{4}y_{ij}^2}$ for the standard halo model, or a sinc function ${\rm sinc} (y_{ij})$ for a delta function approximation of the speed distribution, in these correlation functions, where $y_{ij}$ represents the ratio of pulsar distance $L_{ij}$ with the ALDM de Broglie wavelength $l_c$. This feature distinguishes the ultralight ALDM PTA and PPA detections from the nanoHertz SGWB PTA detection, where the signal spatial correlations matter only at a subleading order.

In Sec.~\ref{sec:PTAanalysis}, we explore statistical properties of the ALDM timing signal and their impacts on the construction of likelihood functions for PTA and combined PTA-PPA Bayesian analyses. We first derive the PDFs for individual timing residuals in some benchmark scenarios. These non-Gaussian PDFs are skewness-free and universally exhibit a sharper peak centered at zero and longer tails, compared to a Gaussian case of the same variance. Despite these features, we show that these PDFs are predominantly determined by their Gaussian components and converge efficiently when their amplitudes are expanded using Gauss-Hermite series, as measured by the Hellinger distance. We are thus highly encouraged to begin with Gaussian approximation for the PTA Bayesian analysis, which greatly simplifies the construction of likelihood function for the signal vector.  
Alternatively, one can construct the likelihood function from the elementary Gaussian variables that compose the ALDM-induced  timing residual. In the case of $\rho_e \ll \rho_p$, we demonstrate that the leading-order signal term for the likelihood function reproduces exactly the likelihood function derived under multivariate Gaussian approximation, in the small signal limit. This validates the use of the multivariate Gaussian distribution for signal vector in~\cite{Luu:2023rgg}, from a different perspective.  
Building on this approximation, we derive the characteristic frequency-domain covariance matrix for calculating the inverse of covariant matrix in the likelihood function. This matrix differs from its SGWB counterpart in both spatial correlation patterns and frequency-space structure.

Finally we expand this study to include the combined PTA-PPA Bayesian analysis. Constructing likelihood function in this case becomes more involved since the leading-order statistical effect for signals arises from their three-point correlation function. Our investigation is mainly for proof of concept. The multivariate Gaussian approximation for the ALDM signals might be still possible in the small-signal limit, but requesting a proper definition of data vector, such that the three-point correlation function can be properly integrated into the likelihood function. Using timing residuals and PA residual quadratics to define the data vector, we show that the three-point correlation function of signal arises from off-diagonal blocks of the full covariant matrix. Alternatively, one can also apply the method of elementary  Gaussian variables to construct the combined PTA-PPA likelihood.
Under the assumption of $\rho_e \ll \rho_p$, together with independent timing and polarization noises, we find that the aforementioned three-point correlation functions emerge as the leading-order cross-correlation signal terms in the marginalized likelihood function, in the small signal limit. 
This finding is significant, laying out the foundation for developing a general Bayesian framework for joint PTA‑PPA analysis.

We thus anticipate the construction of complete likelihoods that account for the non-Gaussian nature of the timing signal in future work, enabling a full Bayesian analysis for both the PTA analysis and the PTA-PPA synergy. One promising direction is to treat the non‑Gaussian statistics in frequency domain, under proper approximations for pulsar arrays. Once such a likelihood is in place, actual inference studies can be carried out to demonstrate the sensitivity potential of this novel search method.

\nocite{footnote1}

\vspace{0.1cm}
\section*{Acknowledgements} 
\vspace{-0.1cm}

We would like to thank Si-Jun Xu for the early collaboration and valuable discussions.
Z.C.C. is supported by the National Natural Science Foundation of China under Grant No.~12405056, the Natural Science Foundation of Hunan Province under Grant No.~2025JJ40006, and the Innovative Research Group of Hunan Province under Grant No. 2024JJ1006.
Q.G.H. is supported by the National Natural Science Foundation of China (Grant No.~12250010). 
J.R. is supported in part by the National Natural Science Foundation of China (Grant No.~12435005).
X.Z. is supported by the National Natural Science Foundation of China (Grant No.~12203004) and by the Fundamental Research Funds for the Central Universities.

\vspace{0.1cm}
\section*{Data Availability} 
\vspace{-0.1cm}

The data that support the findings of this article are openly available~\cite{footnote1}.

\appendix


\section{More Details on the Statistical Properties of ALDM Signals}
\label{app:1}

Let us begin with the statistical properties of the stochastic ALDM field defined in Eq.~(\ref{eq:axionf2}). Consider two variables:
\begin{equation} \label{eq:var}
    x\equiv \alpha \cos\phi,\qquad y\equiv \alpha \sin\phi,
\end{equation}
where $\alpha$ and $\phi$ respect Rayleigh distribution and uniform distribution on $[0,2\pi)$, respectively. The joint distribution of $x$ and $y$ can be calculated through 
\begin{equation}
 p(x,y)\,dx\,dy=\underbrace{\frac \alpha {\sigma^2}e^{-\alpha^2/2\sigma^2}}_{\mathclap{\text{Rayleigh distribution}}}\,d\alpha\cdot\frac 1{2\pi}\,d\phi \, .
\end{equation}
which gives 
\begin{equation}
    p(x,y)= \frac 1{2\pi}\frac \alpha{\sigma^2}e^{-\alpha^2/2\sigma^2}\underbrace{\left[\frac{\partial(x,y)}{\partial(\alpha,\phi)}\right]^{-1}}_{=1/\alpha}
    = \frac{1}{2\pi\sigma^2}e^{-\alpha^2/2\sigma^2}
    = \frac{1}{\sqrt{2\pi\sigma^2}}e^{-x^2/2\sigma^2}\cdot\frac{1}{\sqrt{2\pi\sigma^2}}e^{-y^2/2\sigma^2} \, .
\end{equation}
Therefore, $x$ and $y$ are independent Gaussian variables. 

Generalizing this discussion to the ALDM case, one can define a set of independent Gaussian basis that relies on the ALDM stochastic parameters only: $\{ \alpha_{\mathbf{v}} \cos  \phi _{\mathbf{v}}, \alpha_{\mathbf{v}} \sin \phi _{\mathbf{v}} | \mathbf{v}\in\Omega \}$. The stochastic ALDM field $a(\mathbf{x},t)$ in Eq.~(\ref{eq:axionf2}) can be linearly decomposed in this basis and hence is random Gaussian. This outcome also implies that any linear combinations of the ALDM profiles, including the ALDM-induced PA residual $\Delta \PA^a_{p,n}$ in Eq.~(\ref{eq:DPA1}), should be random Gaussian.

The statistical properties for the ALDM timing signals  have been discussed for single observation in Sec.~\ref{sec:Dtstat}. Below we will offer more details on the case of $\rho_p\approx \rho_e$, where the timing signal can be rewritten as $\dt=2c_1'(X'Y'-U'V')$ (see  Eq.~(\ref{eq:dti3})). $U'$, $V'$, $X'$ and $Y'$ are Gaussian variables with zero mean. $U'$ and $V'$ are correlated, as are $X'$ and $Y'$, but the two sets of variables are independent of each other. The PDF for the composite variables  $x=2c_1'X'Y'$ or $2c_1'U'V'$ have been shown in Eq.~(\ref{eq: 2XY PDF}), which implies (note $r_{X'Y'}=r_{U'V'}$)
\begin{equation} \label{eq:2limits}
    p(x|c'_1)
    \approx \begin{cases}
	\frac{1}{2\pi c_1' \sigma _{\tilde X}\sigma _{\tilde Y}}K_0\left( \frac{\left| x \right|}{2c_1'\sigma _{\tilde X}\sigma _{\tilde Y}} \right) , \quad & r_{\tilde X\tilde Y} \rightarrow 0 \text{\;and\;} x\in \left( -\infty ,\infty \right) \, ; \\
	\frac{1}{2\sqrt{{\pi c_1' \sigma_{\tilde X}^2}x}}\mathrm{e}^{-\frac{x}{4c_1'\sigma_{\tilde X}^{2}}},  & r_{\tilde X\tilde Y}\rightarrow 1 \text{\;and\;} x\in \left[ 0,\infty \right] \, .  \\
\end{cases}  
\end{equation} 
Here we have used $\{\tilde X, \tilde Y\}$ to denote $\{X',Y'\}$  and $\{U',V'\}$. The next step is to determine the PDF for the combination $2c_1'(X'Y'-U'V')$. 

For a random variable 
$x$ with PDF $p(x)$ its characteristic function is defined as 
\begin{equation}  \label{eq: CF}
    C_{x}(\omega)\equiv\left<\mathrm{e}^{\mathrm{i}\omega x}\right>=\int_{-\infty}^{\infty} p(x)\mathrm{e}^{\mathrm{i}\omega x}dx\,.
\end{equation}
This is essentially the inverse Fourier transform of $p(x)$. If $C_X(\omega)$ is well defined at infinity, $p(x)$ can be recovered through Fourier transform
\begin{equation}  \label{eq: Fourier}
    p(x)=\frac{1}{2\pi}\int_{-\infty}^{\infty}C(\omega)\mathrm{e}^{-\mathrm{i}\omega x}d\omega\,.
\end{equation}
The characteristic function $C_{X}(\omega)$ offers a convenient way to study the statistical properties of $x$. For example, the $n^{\mathrm{th}}$ order central moment of $p(x)$ can be calculated as $\left<x^n\right>\equiv\int_{-\infty}^{\infty} p(x) x^n dx =\left.\mathrm{i}^n d^n C_{x}(\omega)/d\omega^n\right|_{\omega=0}$.
Moreover, the characteristic function for difference between two independent and identically distributed variables $x$ and $y$ can be easily derived:  
\begin{equation}
    C_{x-y}(\omega)=C_{x}(\omega)C_{y}(-\omega)\,.
\end{equation}

Applying Eq.~\eqref{eq: CF} to Eq.~\eqref{eq: 2XY PDF} gives the characteristic function of $x=2c_1'X'Y'$ or $2c_1'U'V'$
\begin{equation}  \label{eq: CF_2XY}
  C_{x}(\omega )\equiv \int_{-\infty}^{\infty}{p(x|c'_1)\mathrm{e}^{i \omega x}dx}=\frac{K\pi}{L\sqrt{1-\left( r_{\tilde X\tilde Y}+\mathrm{i}K\omega \right) ^2}}\,,
\end{equation}
with $L\equiv 2\pi c_1'\sigma_{\tilde X}\sigma_{\tilde Y}\sqrt{1-r_{\tilde X\tilde Y}^2}$ and $K\equiv 2c_1'\sigma_{\tilde X}\sigma_{\tilde Y}(1-r_{\tilde X\tilde Y}^2)$. In the two limits discussed in Eq.~(\ref{eq:2limits}), this characteristic function is approximated as 
\begin{equation}
    C_{x}\left( \omega \right) \approx \begin{cases}
	\frac{1}{\sqrt{1+4c_1^2\sigma _{\tilde X}^{2}\sigma _{\tilde Y}^{2}\omega ^2}}, \quad & r_{\tilde X\tilde Y}\rightarrow 0 \text{\;and\;} \omega \in \left( -\infty ,\infty \right)  \, ; \\
	\frac{1}{\sqrt{1-4\mathrm{i}c_1'\sigma _{\tilde X}^2\omega}}, & r_{\tilde X\tilde Y}\rightarrow 1 \text{\;and\;} \omega \in \left( -\infty ,\infty \right)  \, .\\
\end{cases}
\end{equation}
The characteristic function of $\dt$ is then given by  
\begin{eqnarray} \label{eq: CF_2XY-2UV}
\begin{aligned}
C_{\dt}\left( \omega \right) =C_{x}\left( \omega \right) C_{x}\left( -\omega \right) =\frac{\pi ^2}{L^2}\frac{1}{\sqrt{\left( \omega ^2+{K_1}^2 \right) \left( \omega ^2+{K_2}^2 \right)}}\,,
\end{aligned}
\end{eqnarray}
where $K_1\equiv (1+r_{\tilde X\tilde Y})/K$ and  $K_2\equiv (1-r_{\tilde X\tilde Y})/K$. Through  Eq.~\eqref{eq: Fourier}, one can find  
\begin{equation}
    p(\dt|c'_1)=\frac{1}{2\pi}\int_{-\infty}^{\infty} \frac{\pi^2}{L^2}\frac{1} 
    {\sqrt{\left( \omega ^2+{K_1}^2 \right) \left( \omega ^2+{K_2}^2 \right)}}\mathrm{e}^{-\mathrm{i}\omega \dt}d\omega \,.
\end{equation}
While analytically evaluating this integral is challenging, the $\dt$ PDF has relatively simple forms in the two special limits, as shown in Eq.~\eqref{eq:fDtlimit}, 
\begin{eqnarray}
     && r_{\tilde X\tilde Y}=1:\; p(\delta t|c'_1)=\frac{1}{\pi\sigma}K_0\left( \frac{\left| \dt \right|}{\sigma} \right) \, , \quad  \sigma=4 c_1'\sigma_{\tilde X}\sigma_{\tilde Y}   \, ; \\
     && r_{\tilde X\tilde Y}=0:\; p(\delta t|c'_1)=\frac{1}{2b}\mathrm{e}^{-\frac{\left| \dt \right|}{b}} \, ,  \quad  \quad  \quad  \quad  b=2c_1'\sigma_{\tilde X}\sigma_{\tilde Y}\, .
\end{eqnarray}
For $0<r_{\tilde X\tilde Y}<1$, the $\dt$ PDF is sandwiched between these two limits, and could be numerically calculated.

\section{More Details on Derivations Involving Elementary Gaussian Variables}
\label{app:2}

In this Appendix, we offer a more detailed explanation of the calculations for several key formulas discussed in the main text. These derivations leverage the elementary Gaussian variables that constitute the ALDM timing and polarization signals, significantly simplifying the derivation of the correlation functions in Sec.~\ref{sec:correlation} and the marginalized likelihoods in Sec.~\ref{sec:PTAanalysis}.

To begin, the ALDM timing and polarization signals $\bDt^a$ and $\bDPA^a$ can be expressed in a compact form (i.e. the component form given in Eqs.~(\ref{eq:DPA1c}) and (\ref{eq:DtXY})), 
\begin{eqnarray}\label{eq:bDtbDPAapp}
\bDt^a = \frac{\pi G}{2m_{a}^{3}}\sum_{i=0,1}{\left( -1 \right) ^i\rho( \mathbf{x}_{p}^{\left( i \right)}) \bX^{(i)}\odot\bY^{(i)}},\quad
\bDPA^a =-\frac{g_{a\gamma\gamma}}{m_a}\sum_{i=0,1}(-1)^i \sqrt{\rho(\mathbf{x}_p^{(i)})}\bX^{(i)}\,,
\end{eqnarray}
where $i=0$ and 1 denote the ``Earth'' and ``pulsar'' terms, respectively. $\bX^{(i)}=(X^{(i)}_{1,1},..., X^{(i)}_{p,n},...,X^{(i)}_{\mathcal{N},N_\mathcal{N}})^T$ and $\bY^{(i)}=(Y^{(i)}_{1,1},..., Y^{(i)}_{p,n},...,Y^{(i)}_{\mathcal{N},N_\mathcal{N}})^T$ represent the four sets of elementary Gaussian variables, with their components given by (i.e. Eqs.~(\ref{eq:Xpn}) and (\ref{eq:Ypn})) 
\begin{eqnarray}\label{eq:XYpn}
    X_{p,n}^{(i)}\equiv  \sum_{\mathbf{v}\in \Omega}\mC_{\mathbf{v}}\cos\Big[\vartheta_{\mathbf{v}}(\mathbf{x}_p^{(i)},t_{p,n}^{(i)})\Big],\quad
    Y_{p,n}^{(i)}\equiv  \sum_{\mathbf{v}\in \Omega}\mC_{\mathbf{v}}\sin\Big[\vartheta_{\mathbf{v}}(\mathbf{x}_p^{(i)},t_{p,n}^{(i)})\Big]\,,
\end{eqnarray}
where $\mC_{\mathbf{v}}\equiv (\Delta v ) ^{3/2}\alpha_{\mathbf{v}}\sqrt{f(\mathbf{v})}$ denotes the amplitude and $\vartheta_{\mathbf{v}}(\mathbf x,t)\equiv m_a(t-\mathbf{v}\cdot \mathbf{x})+\phi_{\mathbf{v}}$ represents the phase. Here,  $\alpha_{\mathbf{v}}$ and $\phi_{\mathbf{v}}$ are independent random variables defined on the phase space lattice sites $\mathbf{v}\in\Omega$, following Rayleigh and uniform distributions, respectively.

These elementary Gaussian variables have zero mean, and  their statistical properties are fully determined by their two-point functions. Specifically, for the variable $\bX^{(i)}$, the covariance matrix $\bC_X^{(ij)}=\langle\bX^{(i)}(\bX^{(j)})^T\rangle$ is symmetric both for a given set of $i$ and $j$, and with respect to interchanging $i$ and $j$. Its component $(\bC_X^{(ij)})_{pn,qm}$ (i.e. Eq.~(\ref{eq:bCXij})) can be derived as
\begin{eqnarray}\label{eq:X2pt}
  \langle X^{(i)}_{p,n}X^{(j)}_{q,m}\rangle  &=&\sum_{\mathbf{v},\mathbf{v}'}\left\langle \mC_{\mathbf{v}}\mC_{\mathbf{v}'}\right\rangle\left\langle\cos\Big[\vartheta_{\mathbf{v}}(\mathbf{x}_p^{(i)},t_{p,n}^{(i)})\Big]\cos\Big[\vartheta_{\mathbf{v}'}(\mathbf{x}_q^{(j)},t_{q,m}^{(j)})\Big]\right\rangle\nonumber\\
  &=&\int d^3\mathbf{v}f(\mathbf{v})\cos \Big[ m_a ( t_{p,n}^{\left( i \right)}-t_{q,m}^{\left( j \right)}) -m_a\mathbf{v}\cdot \mathbf{x}_{pq}^{\left( ij \right)} \Big]  \nonumber \\
  &=& e^{-\frac{1}{4}(y_{pq}^{ij})^2} \cos \Big[ m_a ( t_{p,n}^{\left( i \right)}-t_{q,m}^{\left( j \right)}) +m_a\mathbf{v}_{\odot}\cdot \mathbf{x}_{pq}^{\left( ij \right)} \Big]\,, 
\end{eqnarray}
with $\mathbf{x}_{pq}^{(ij)}\equiv \mathbf{x}_p^{(i)}-\mathbf{x}_q^{(j)}$,  $y^{ij}_{pq}\equiv |\mathbf{x}^{(ij)}_{pq}|/l_c$ and $l_c=1/(m_a v_0)$. Here, $\langle\cos(a+\phi_{\mathbf{v}})\cos(b+\phi_{\mathbf{v}'})\rangle=\frac{1}{2}\cos(a-b)\delta_{\mathbf{v},\mathbf{v}'}$ and $\langle\alpha_{\mathbf{v}}\alpha_{\mathbf{v'}}\rangle=2\delta_{\mathbf{v,v'}}$, based on the respective distributions for $\phi_{\mathbf{v}}$ and $\alpha_{\mathbf{v}}$. 
For the phase space integration, we find 
$\int d^3\mathbf{v}f(\mathbf{v})\cos(a-\mathbf{v}\cdot \mathbf{z})=\cos(a+\mathbf{v}_{\odot}\cdot\mathbf{z}) e^{-\frac{1}{4}(v_0|\mathbf{z}|)^2}$ with the SHM velocity distribution given in Eq.~(\ref{eq: SHM}). Applying the same strategy to the variable $\bY^{(i)}$, we have  
\begin{eqnarray}\label{eq:Y2pt}
  \langle Y^{(i)}_{p,n}Y^{(j)}_{q,m}\rangle  &=&\sum_{\mathbf{v},\mathbf{v}'}\left\langle \mC_{\mathbf{v}}\mC_{\mathbf{v}'}\right\rangle\left\langle\sin\Big[\vartheta_{\mathbf{v}}(\mathbf{x}_p^{(i)},t_{p,n}^{(i)})\Big]\sin\Big[\vartheta_{\mathbf{v}'}(\mathbf{x}_q^{(j)},t_{q,m}^{(j)})\Big]\right\rangle\nonumber\\
  &=&\int d^3\mathbf{v}f(\mathbf{v})\cos \Big[ m_a ( t_{p,n}^{\left( i \right)}-t_{q,m}^{\left( j \right)}) -m_a\mathbf{v}\cdot \mathbf{x}_{pq}^{\left( ij \right)} \Big]  \nonumber \\
  &=& e^{-\frac{1}{4}(y_{pq}^{ij})^2} \cos \Big[ m_a ( t_{p,n}^{\left( i \right)}-t_{q,m}^{\left( j \right)}) +m_a\mathbf{v}_{\odot}\cdot \mathbf{x}_{pq}^{\left( ij \right)} \Big]\,, 
\end{eqnarray}
which aligns with $\langle X^{(i)}_{p,n}X^{(j)}_{q,m}\rangle$ in Eq.~(\ref{eq:X2pt}). 
Consequently, the covariance matrix for $\bY^{(i)}$ is the same as that for $\bX^{(i)}$, i.e. $\langle\bY^{(i)}(\bY^{(j)})^T\rangle=\langle\bX^{(i)}(\bX^{(j)})^T\rangle=\bC_X^{(ij)}$.
For their cross terms, i.e. $\bC_{XY}^{(ij)}=\langle\bX^{(i)}(\bY^{(j)})^T\rangle$, we calculate 
\begin{eqnarray}\label{eq:XY2pt}
  \langle X^{(i)}_{p,n}Y^{(j)}_{q,m}\rangle  &=&\sum_{\mathbf{v},\mathbf{v}'}\left\langle \mC_{\mathbf{v}}\mC_{\mathbf{v}'}\right\rangle\left\langle\cos\Big[\vartheta_{\mathbf{v}}(\mathbf{x}_p^{(i)},t_{p,n}^{(i)})\Big]\sin\Big[\vartheta_{\mathbf{v}'}(\mathbf{x}_q^{(j)},t_{q,m}^{(j)})\Big]\right\rangle\nonumber\\
  &=&-\int d^3\mathbf{v}f(\mathbf{v})\sin \Big[ m_a ( t_{p,n}^{\left( i \right)}-t_{q,m}^{\left( j \right)}) -m_a\mathbf{v}\cdot \mathbf{x}_{pq}^{\left( ij \right)} \Big]  \nonumber \\
  &=& - e^{-\frac{1}{4}(y_{pq}^{ij})^2} \sin \Big[ m_a ( t_{p,n}^{\left( i \right)}-t_{q,m}^{\left( j \right)}) +m_a\mathbf{v}_{\odot}\cdot \mathbf{x}_{pq}^{\left( ij \right)} \Big]\,, 
\end{eqnarray}
using $\int d^3\mathbf{v}f(\mathbf{v})\sin(a-\mathbf{v}\cdot \mathbf{z})=\sin(a+\mathbf{v}_{\odot}\cdot\mathbf{z}) e^{-\frac{1}{4}(v_0|\mathbf{z}|)^2}$. This leads to $(\bC_{XY}^{(ij)})_{pn,qm}$ in Eq.~(\ref{eq:bCXYij}). 
Given that $\langle X^{(j)}_{q,m}Y^{(i)}_{p,n}\rangle=-\langle X^{(i)}_{p,n}Y^{(j)}_{q,m}\rangle $, the matrix $\bC_{XY}^{(ij)}$ is antisymmetric, indicating that  $\langle X_{p,n}^{(i)}Y_{p,n}^{(i)} \rangle=0$.   

With the information from Eqs.~(\ref{eq:X2pt})-(\ref{eq:XY2pt}), the correlation functions of ALDM signals calculated in Sec.~\ref{sec:correlation} can be readily obtained. Specifically, the two-point functions for the polarization signal in Eq.~(\ref{eq:DPAcorr1}) can be found as
\begin{eqnarray}
&&\langle \dPA_{p,n}^{a} \dPA_{q,m}^{a}\rangle \nonumber\\ 
    &=& \frac{g^2_{a\gamma\gamma}}{m^2_a}\sum_{i,j}(-1)^{i+j} \sqrt{\rho(\mathbf{x}_p^{(i)})\rho(\mathbf{x}_q^{(j)})}\, \langle X_{p,n}^{(i)}X_{q,m}^{(i)}\rangle \nonumber\\
    &=& \frac{g_{a\gamma\gamma}^2}{m_a^2}\sum_{i,j}(-1)^{i+j}\sqrt{\rho(\mathbf{x}_p^{(i)})\rho(\mathbf{x}_q^{(j)})}e^{-\frac{1}{4}(y_{pq}^{ij})^2} \cos \Big[ m_a \left( t_{p,n}^{\left( i \right)}-t_{q,m}^{\left( j \right)}\right) +m_a\mathbf{v}_{\odot}\cdot \mathbf{x}_{pq}^{\left( ij \right)} \Big]\,.
\end{eqnarray}
The corresponding matrix form is then given by
\begin{eqnarray}\label{eq:CPAaapp}
    \bC_{\rm PA}^{a}=\langle \bDPA^a (\bDPA^a)^T\rangle
    =\frac{g_{a\gamma\gamma}^2}{m_a^2}\sum_{i,j}(-1)^{i+j}\sqrt{\rho(\mathbf{x}_p^{(i)})\rho(\mathbf{x}_q^{(j)})}\, \bC_{X}^{(ij)} \, ,  
\end{eqnarray}
showing the linear dependence of the covariance matrix of $\bDPA^a$ on those of the elementary Gaussian variables $\bX^{(i)}$.

The two-point functions for the timing signal in Eq.~(\ref{eq:Dtcorr}) appear more complicated due to the quadratic field dependence and can be derived as 
\begin{eqnarray}     
&&\left< \dt_{p,n}^{a}\dt_{q,m}^{a} \right>\nonumber\\
&=& \frac{\pi ^2G^2}{4m_{a}^{6}}\sum_{i,j}{\left( -1 \right) ^{i+j}}\rho( \mathbf{x}_{p}^{\left( i \right)}) \rho(\mathbf{x}_{q}^{\left( j \right)})
\left\langle X_{p,n}^{(i)} Y_{p,n}^{(i)} X_{q,m}^{(j)} Y_{q,m}^{(j)} \right\rangle \nonumber\\
&=&\frac{\pi ^2G^2}{4m_{a}^{6}}\sum_{i,j}{\left( -1 \right) ^{i+j}}\rho( \mathbf{x}_{p}^{\left( i \right)}) \rho(\mathbf{x}_{q}^{\left( j \right)})
\Big[\langle X_{p,n}^{(i)}X_{q,m}^{(j)} \rangle \langle Y_{p,n}^{(i)}  Y_{q,m}^{(j)} \rangle +\langle X_{p,n}^{(i)} Y_{q,m}^{(j)} \rangle \langle Y_{p,n}^{(i)}  X_{q,m}^{(j)}  \rangle  \Big]\nonumber\\
&=&\frac{\pi ^2G^2}{4m_{a}^{6}}\sum_{i,j}{\left( -1 \right) ^{i+j}}\rho( \mathbf{x}_{p}^{\left( i \right)}) \rho(\mathbf{x}_{q}^{\left( j \right)})
\Big[(\bC_X^{(ij)})_{pn,qm}(\bC_X^{(ij)})_{pn,qm} +(\bC_{XY}^{(ij)})_{pn,qm}  (\bC_{XY}^{(ji)})_{qm,pn} \Big]\nonumber\\
&=&\frac{\pi ^2G^2}{4m_{a}^{6}}\sum_{i,j}{\left( -1 \right) ^{i+j}\rho ( \mathbf{x}_{p}^{\left( i \right)}) \rho(\mathbf{x}_{q}^{\left( j \right)})}e^{-\frac{1}{2}\left( y_{pq}^{ij} \right) ^2}\cos \Big[ 2m_a\left( t_{p,n}^{(i)}-t_{q,m}^{(j)} \right) +2m_a\mathbf{v}_{\odot}\cdot  \mathbf{x}_{pq}^{(ij)} \Big] \,. 
\end{eqnarray}
Unlike the previous case, this derivation  
is fundamentally related to the four-point correlation functions of the elementary Gaussian variables, which can be decomposed into two distinct sets of combinations of their two-point functions, given that $\langle X_{p,n}^{(i)}Y_{p,n}^{(i)} \rangle=0$. The matrix form then becomes 
\begin{eqnarray}\label{eq:Ctafullapp}
  \bC_{t}^{a}=\langle \bDt^a (\bDt^a)^T\rangle
    =\frac{\pi^2 G^2}{4m_a^6}\sum_{i,j}(-1)^{i+j}\rho ( \mathbf{x}_{p}^{\left( i \right)}) \rho(\mathbf{x}_{q}^{\left( j \right)})\left[\bC_{X}^{(ij)}\odot \bC_{X}^{(ij)}- \bC_{XY}^{(ij)}\odot\bC_{XY}^{(ij)}\right]\,,
\end{eqnarray}
illustrating the quadratic dependence of the covariance matrix of $\bDt^a$ on those of $\bX^{(i)}$ and $\bY^{(i)}$.

The three-point functions for the polarization and timing signals in Eq.~(\ref{eq:DPADPADtcorr}) can be similarly derived, resulting in
\begin{eqnarray} 
    && \langle \dPA^a_{p,n}\dPA^a_{q,m}\dt^a_{r,l}\rangle \nonumber \\
    &=&
    \frac{\pi G g^2_{a\gamma\gamma}}{2m_a^5} \sum_{i,j,k}(-1)^{i+j+k}\sqrt{\rho(\mathbf{x}_p^{(i)})\rho(\mathbf{x}_q^{(j)})}\rho(\mathbf{x}_r^{(k)})\left\langle X_{p,n}^{(i)}X_{q,m}^{(j)}X_{r,l}^{(k)}Y_{r,l}^{(k)}\right\rangle\nonumber\\
    &=&\frac{\pi G g^2_{a\gamma\gamma}}{2m_a^5} \sum_{i,j,k}(-1)^{i+j+k}\sqrt{\rho(\mathbf{x}_p^{(i)})\rho(\mathbf{x}_q^{(j)})}\rho(\mathbf{x}_r^{(k)})\left[\left\langle X_{p,n}^{(i)}X_{r,l}^{(k)}\right\rangle\left\langle X_{q,m}^{(j)}Y_{r,l}^{(k)}\right\rangle+\left\langle X_{q,m}^{(j)}X_{r,l}^{(k)}\right\rangle\left\langle X_{p,n}^{(i)}Y_{r,l}^{(k)}\right\rangle\right]\nonumber\\
    &=&\frac{\pi G g^2_{a\gamma\gamma}}{2m_a^5} \sum_{i,j,k}(-1)^{i+j+k}\sqrt{\rho(\mathbf{x}_p^{(i)})\rho(\mathbf{x}_q^{(j)})}\rho(\mathbf{x}_r^{(k)})\left[(\bC_{X}^{(ik)})_{pn,rl}(\bC_{XY}^{(jk)})_{qm,rl}+(\bC_{X}^{(jk)})_{qm,rl}(\bC_{XY}^{(ik)})_{pn,rl}\right]\nonumber\\
    &=&-\frac{\pi Gg_{a\gamma \gamma}^{2}}{2m_{a}^{5}}\sum_{i,j,k}{\left( -1 \right) ^{i+j+k}}\sqrt{\rho(\mathbf{x}_p^{(i)})\rho(\mathbf{x}_q^{(j)})}\rho(\mathbf{x}_r^{(k)})\, e^{-\frac{1}{4}\left( y_{pr}^{ik} \right) ^2} e^{-\frac{1}{4}\left( y_{qr}^{jk} \right) ^2 }\nonumber\\
    && \quad\quad\quad\quad\quad\quad \times \sin \Big[ m_a( t_{p,n}^{\left( i \right)}+t_{q,m}^{\left( j \right)}-2t_{r,l}^{\left( k \right)}) +m_a\mathbf{v}_{\odot}\cdot (\mathbf{x}_{pr}^{\left( ik \right)}+ \mathbf{x}_{qr}^{\left( jk \right)}) \Big]\,.
\end{eqnarray} 
Once again, this is related to the four-point correlation functions of the elementary Gaussian variables, which can be decomposed into two distinct sets of combinations for their two-point functions.
Using this framework, the connections between these three types of correlation functions become quite straightforward. In particular, the similarity between $\left< \dt_{p,n}^{a}\dt_{q,m}^{a} \right>$ and $\langle \dPA^a_{p,n}\dPA^a_{q,m}\dt^a_{r,l}\rangle$ is especially evident.


In Sec.~\ref{sec:PTAanalysis}, we construct likelihood functions for PTA and combined PTA-PPA analyses of ALDM signals, considering the non-Gaussian nature of the timing signal.  For both analyses, a general method for deriving the marginalized likelihood 
involves marginalizing over its elementary Gaussian variables in the small-signal limit. This approach explicitly illustrates the connection of marginalized likelihood to the characteristic correlation functions discussed in Sec.~\ref{sec:correlation}.

To demonstrate this general method, let us consider the scenario where $\rho_p\gg\rho_e$. In this limit, the ALDM signal vectors in Eq.~(\ref{eq:bDtbDPAapp}) reduce to
\begin{eqnarray}
\bDt^a=c_1 \bD_X \bY, \quad 
\bDPA^a=c_2 \bX\,,
\end{eqnarray}
where $\bX$ and $\bY$ are shorthand for $\bX^{(1)}$ and $\bY^{(1)}$, respectively. Here, $\bD_X$ represents $\bX$ in matrix form, {\it i.e.},  $\bD_X={\rm diag}(\bX)$. The two dimensionless coefficients are defined as $c_1 = \pi G\rho_p/(2m_a^3)$ and    $c_2=g_{a\gamma\gamma}\sqrt{\rho_p}/m_a$. With the two-point correlation functions of $\bX$ and $\bY$ in Eqs.~(\ref{eq:X2pt})-(\ref{eq:XY2pt}), their joint multivariate Gaussian PDF is then given by
\begin{eqnarray}\label{eq:fXY}
p(\{\bX,\bY\}|\bDvth^a)
&=&c_{XY}
\exp\left[-\frac{1}{2}\begin{pmatrix}
  \bX^T & \bY^T  
\end{pmatrix}\begin{pmatrix}
  \bC_X & \bC_{XY} \\
  -\bC_{XY} & \bC_X
\end{pmatrix}^{-1}\begin{pmatrix}
  \bX  \\
  \bY 
\end{pmatrix}\right]\nonumber\\
&=&c_{XY}
\exp\left[-\frac{1}{2} \left(\bX^T \bS^{-1} \bX+\bY^T \bS^{-1} \bY+\bY^T \bS^{-1}\bC_{XY}\bC_X^{-1} \bX-\bX^T \bC_X^{-1} \bC_{XY}  \bS^{-1} \bY\right)\right],\nonumber\\
\end{eqnarray}
where $\bC_X$ and $\bC_{XY}$ are shorthand for $\bC_X^{(11)}$ and $\bC_{XY}^{(11)}$, and $\bS=\bC_X+\bC_{XY}\bC_X^{-1}\bC_{XY}$ is derived from the inversion of the $2\times 2$ block covariance matrix in the first line. The prefactor is related to the determinant of covariance matrix, i.e. $c_{XY}=[\det(2\pi (\bC_X+i \bC_{XY}))\det(2\pi (\bC_X-i \bC_{XY}))]^{-1/2}$.

For the PTA analysis, 
the marginalized likelihood $p(\bDt|\bDvth_t)$ can be derived by integrating over the elementary Gaussian variables that make up $\bDt^a$, i.e. Eq.~(\ref{eq:eq:LmDtXY}). 
By substituting the explicit form of the joint PDF in Eq.~(\ref{eq:fXY}), the marginalized likelihood is then expressed as 
\begin{eqnarray}\label{eq:LmDtvec1p2}
p(\bDt|\bDvth_t)
&=& \frac 1{\sqrt{\det(2\pi\boldsymbol C^{n}_t)}} \int\exp\left[-\frac{1}{2}\big(\bDt_o-c_1 \bD_X \bY\big)^T(\bC^{n}_t)^{-1}\big(\bDt_o-c_1 \bD_X \bY\big)\right]p(\{\bX,\bY\}|\bDvth^a) d\bX d\bY\nonumber\\
&=& \frac{c_{XY}}{\sqrt{\det(2\pi\boldsymbol C^{n}_t)}} \int\exp\left[-\frac{1}{2}\big(\bDt_o-c_1 \bD_X \bY\big)^T(\bC^{n}_t)^{-1}\big(\bDt_o-c_1 \bD_X \bY\big)\right]
\exp\left[-\frac{1}{2} \bX^T \bS^{-1} \bX\right]\nonumber\\
&& \times \exp\left[-\frac{1}{2} \bY^T \bS^{-1} \bY\right] \exp\left[\frac{1}{2} (-\bY^T \bS^{-1}\bC_{XY}\bC_X^{-1} \bX+\bX^T \bC_X^{-1} \bC_{XY}  \bS^{-1} \bY)\right] d\bX d\bY  \nonumber\\
&=& \frac{c_{XY}} {\sqrt{\det(2\pi\boldsymbol C^{n}_t)}}  \exp\left[-\frac{1}{2}\bDt_o^T(\bC^{n}_t)^{-1}\bDt_o \right] \frac{1}{\sqrt{\det(2\pi \boldsymbol{A})}}  \nonumber\\
&&\int  \exp\left[-\frac{1}{2}\bX^T(\bS^{-1}-\bB \boldsymbol{A}^{-1}\bB) \bX\right] d\bX\,.
\end{eqnarray}
In the last step, we carry out the standard Gaussian integral over $\bY$, where $\boldsymbol{A}=\bS^{-1}+c_1^2\bD_{X} (\bC^{n}_t)^{-1}\bD_{X}$ and $\bB=c_1 \bD_t-\bS^{-1}\bC_{XY}\bC_X^{-1}$ with $\bD_t={\rm diag}((\bC_t^{n})^{-1}\bDt_o)$.

To marginalize over $\bX$, we consider the small-signal limit, where  $c_1^2\Vert(\bC^{n}_t)^{-1}\Vert$ and $\Vert c_1 \bD_t\Vert\ll1$, and expand $p(\bDt|\bDvth_t)$ w.r.t. $c_1$:   
\begin{eqnarray}\label{eq:LmDtvec1}
p(\bDt|\bDvth_t)
&\propto& \exp\left[-\frac{1}{2}\bDt_o^T(\bC^{n}_t)^{-1}\bDt_o \right] \int  d\bX 
    \exp\bigg[-\frac{1}{2} \bX^T \left(\bS^{-1}-\bB \bS \bB\right) \bX\bigg]\nonumber\\
&\propto& \exp\left[-\frac{1}{2}\bDt_o^T(\bC^{n}_t)^{-1}\bDt_o \right] \det(\bS^{-1}-\bB \bS \bB)^{-1/2}\,.
\end{eqnarray}
Here, to compare with the Gaussian approximation $p^{(g)}(\bDt|\bDvth_t)$ in Eq.~(\ref{eq:LmGaussian}), we focus on the exponential term that is quadratic in $\bDt_o$ for demonstration purposes. As a result, only the leading term in $\boldsymbol{A}^{-1}$ in the exponential expression of Eq.~(\ref{eq:LmDtvec1p2}) is retained. The matrix within the determinant can be further simplified: 
\begin{eqnarray}
    \bS^{-1}-\bB \bS \bB=\bC_X^{-1}+c_1\Big(\bD_t\bC_{XY}\bC_X^{-1}-\bC_X^{-1}\bC_{XY}\bD_t\Big)-c_1^2\bD_t\bS\bD_t\,,
\end{eqnarray}
given $\bS^{-1}=\bC_X^{-1}-\bC_X^{-1}\bC_{XY}\bS^{-1}\bC_{XY}\bC_X^{-1}$. By perturbatively expanding this  determinant, we finally obtain 
\begin{eqnarray}
p(\bDt|\bDvth_t)
    &\propto& \exp\left[-\frac{1}{2}\bDt_o^T(\bC^{n}_t)^{-1}\bDt_o \right]\bigg[1+\frac{1}{2}c_1^2\tr(\bC_X \bD_t\bS \bD_t) + \frac{1}{2}c_1^2\Big(\tr(\bC_{XY}\bD_t\bC_{XY}\bD_t)\nonumber\\
    &&-\tr(\bC_X\bD_t\bC_{XY}\bC_X^{-1}\bC_{XY}\bD_t)\Big) +\mathcal{O}(c_1^3)\bigg]\nonumber\\
    &\propto& \exp\left[-\frac{1}{2}\bDt_o^T(\bC^{n}_t)^{-1}\bDt_o \right] \bigg[1+\frac{1}{2}c_1^2\Big(\tr(\bC_X \bD_t\bC_X \bD_t)+\tr(\bC_{XY}\bD_t\bC_{XY}\bD_t)\Big)+\mathcal{O}(c_1^3)\bigg]\nonumber\\
    &\propto& \exp\left[-\frac{1}{2}\bDt_o^T(\bC^{n}_t)^{-1}\bDt_o \right] \bigg[1+\frac{1}{2}c_1^2\bDt_o^T(\bC^{n}_t)^{-1}\Big(\bC_X\odot\bC_X-\bC_{XY}\odot\bC_{XY}\Big)(\bC^{n}_t)^{-1}\bDt_o\bigg]\nonumber\\
    &\propto& \exp\left[-\frac{1}{2}\bDt_o^T(\bC^{n}_t)^{-1}\bDt_o \right] \bigg[1+\frac{1}{2}\bDt_o^T(\bC^{n}_t)^{-1}\bC_t^{a}(\bC^{n}_t)^{-1}\bDt_o +\mathcal{O}(c_1^3)\bigg]\, ,
\end{eqnarray}
where we have used the signal covariance matrix $\bC_t^{a}$ in Eq.~(\ref{eq:Ctafullapp}) for the $\rho_p\gg\rho_e$ case in the last line. 
This gives rise to Eq.~(\ref{eq:LmDtvec2}), which aligns perfectly with the small-signal limit $p^{(g)}$ in Eq.~(\ref{eq:LmGaussianexp}).

For the combined PTA-PPA analysis, assuming first that the random noises for the timing and polarization data are independent, the marginalized likelihood  can be derived by integrating $p(\bV|\bDvth^n,\bV^a)$ from Eq.~(\ref{eq:LDtDPA}) over the two elementary Gaussian variables
\begin{eqnarray}\label{eq:LmDtvec1p2}
p(\bV|\bDvth)
&=& \frac {1}{\sqrt{\det(2\pi \bC^{n}_t)\det(2\pi \bC^{n}_{\PA})}} \int\exp\left[-\frac{1}{2}\big(\bDt_o-c_1 \bD_X \bY\big)^T(\bC^{n}_t)^{-1}\big(\bDt_o-c_1 \bD_X \bY\big)\right]\nonumber\\
&&\times \exp\left[-\frac{1}{2}\big(\bDPA_o-c_2 \bX\big)^T(\bC^{n}_{\PA})^{-1}\big(\bDPA_o-c_2 \bX\big)\right]p(\{\bX,\bY\}|\bDvth^a) d\bX d\bY \,.
\end{eqnarray}
Following the same approach, we find the marginalized likelihood in the small-signal limit: 
\begin{eqnarray}
p(\bV|\bDvth)
&\propto&\int d\bX\,d\bY\exp\left[-\frac{1}{2}\big(\bDt_o-c_1 \bD_X \bY\big)^T(\bC^{n}_t)^{-1}\big(\bDt_o-c_1 \bD_X \bY\big)\right]\nonumber\\
&& \times \exp\left[-\frac{1}{2}\big(\bDPA_o-c_2 \bX\big)^T(\bC^{n}_{\PA})^{-1}\big(\bDPA_o-c_2 \bX\big)\right]\nonumber\\
&& \times \exp\left[-\frac{1}{2} \left(\bX^T \bS^{-1} \bX+\bY^T \bS^{-1} \bY+\bY^T \bS^{-1}\bC_{XY}\bC_X^{-1} \bX-\bX^T \bC_X^{-1} \bC_{XY}  \bS^{-1} \bY\right)\right] \nonumber\\
&\propto& \exp\left[-\frac{1}{2}\left(\bDt_o^T(\bC_{t}^{n})^{-1}\bDt_o+\bDPA_o^T(\bC^{n}_{\PA})^{-1}\bDPA_o\right) \right]\int  d\bX \nonumber\\
&&\exp\bigg[-\frac{1}{2} \bX^T \left(c_2^2(\bC^{n}_{\PA})^{-1}+\bS^{-1}-\bB \bS \bB\right) \bX
    +c_2\left(\bX^T(\bC^{n}_{\PA})^{-1}\bDPA_o+\bDPA_o^T(\bC^{n}_{\PA})^{-1}\bX\right) \bigg]\nonumber\\
    &\propto& \exp\left[-\frac{1}{2}\left(\bDt_o^T(\bC_{t}^{n})^{-1}\bDt_o+\bDPA_o^T(\bC^{n}_{\PA})^{-1}\bDPA_o\right) \right]\det\Big(\bS^{-1}-\bB\bS\bB+c_2^2(\bC_{\PA}^n)^{-1}\Big)^{-1/2}\nonumber\\
    && \times \exp\bigg[\frac{1}{2}c_2^2\bDPA_o^T(\bC^{n}_{\PA})^{-1}\Big(\bS^{-1}-\bB \bS \bB+c_2^2(\bC^{n}_{\PA})^{-1}\Big)^{-1}(\bC^{n}_{\PA})^{-1}\bDPA_o\bigg]  \, .
\end{eqnarray}
The exponential factor in the last line captures the cross-correlation between the timing and polarization signals. Considering that  
\begin{eqnarray}
\Big(\bS^{-1}-\bB \bS \bB+c_2^2(\bC^{n}_{\PA})^{-1}\Big)^{-1}
    \approx \bC_X-c_1(\bC_X\bD_t\bC_{XY}-\bC_{XY}\bD_t\bC_X)+\mathcal{O}(c_1^2)+\mathcal{O}(c_2^2),
\end{eqnarray}
the marginalized likelihood is  finally given by
\begin{eqnarray}
p(\bV|\bDvth)
&\propto&
\exp\left[-\frac{1}{2}\left(\bDt_o^T(\bC_{t}^{n})^{-1}\bDt_o+\bDPA_o^T(\bC^{n}_{\PA})^{-1}\bDPA_o\right) \right]\nonumber\\
    && \times \exp\bigg[\frac{1}{2}\bDPA_o^T(\bC^{n}_{\PA})^{-1}\bC^a_\PA(\bC^{n}_{\PA})^{-1}\bDPA_o\bigg]\bigg[1+\frac{1}{2}\bDt_o^T(\bC^{n}_t)^{-1}\bC_t^{a}(\bC^{n}_t)^{-1}\bDt_o\bigg]\nonumber\\
    && \times \exp\bigg[\frac{1}{2}c_1c_2^2\bDPA_o^T(\bC^{n}_{\PA})^{-1}(\bC_X\bD_t\bC_{XY}-\bC_{XY}\bD_t\bC_X)(\bC^{n}_{\PA})^{-1}\bDPA_o\bigg] \, .
\end{eqnarray}
where we have used the signal covariance matrix $\bC_\PA^{a}=c_2^2\bC_X$ in Eq.~(\ref{eq:CPAaapp}) for the $\rho_p\gg\rho_e$ case. This results in Eq.~(\ref{eq:TPL}), which includes information about the two-point correlations of \(\bDPA^a\) and \(\bDt^a\), as well as their three-point cross-correlation functions, as detailed in the main text.

\clearpage
\newpage

\bibliography{reference}

\end{document}